\documentclass[lettersize,journal]{IEEEtran}
\usepackage{amsmath,amsfonts}
\usepackage{xcolor}
\usepackage{algorithmicx}
\usepackage{algorithm}
\usepackage{array}
\usepackage[caption=false,font=normalsize,labelfont=sf,textfont=sf]{subfig}
\usepackage{textcomp}
\usepackage{stfloats}
\usepackage{url}
\usepackage{verbatim}
\usepackage{graphicx}
\usepackage{cite}
\usepackage{float}
\usepackage{caption}
\usepackage{subcaption}
\usepackage{color}
\usepackage{algpseudocode}
\usepackage{listings}
\usepackage{booktabs}
\usepackage{multirow}
\usepackage{makecell}
\usepackage[normalem]{ulem}
\lstdefinestyle{json}{
  basicstyle=\ttfamily\small,
  literate=
   *{0}{{0}}{1}
    {1}{{1}}{1}
    {2}{{2}}{1}
    {3}{{3}}{1}
    {4}{{4}}{1}
    {5}{{5}}{1}
    {6}{{6}}{1}
    {7}{{7}}{1}
    {8}{{8}}{1}
    {9}{{9}}{1}
    {,}{{,}}{1}
    {:}{{:}}{1}
    {\{}{{\{}}{1}
    {\}}{{\}}}{1}
    {[}{{[}}{1}
    {]}{{]}}{1},
}

\floatname{algorithm}{}


\floatstyle{boxed}
\restylefloat{algorithm}

\begin{document} 
\title{NaturalVoices: A Large-Scale, Spontaneous and Emotional Podcast Dataset for Voice Conversion} 

\author{Zongyang Du, Shreeram Suresh Chandra, Ismail Rasim Ulgen, Aurosweta Mahapatra ~\IEEEmembership{Student~Member,~IEEE}, \\ Ali N. Salman~\IEEEmembership{Member, IEEE}, Carlos Busso~\IEEEmembership{Fellow,~IEEE,}  Berrak~Sisman~\IEEEmembership{Senior~Member,~IEEE}
\thanks{Z. Du is with the Department of Electrical and Computer Engineering, University of Texas at Dallas, Richardson, TX 75080 USA, and also with the Department of Electrical and Computer Engineering, Johns Hopkins University, Baltimore, MD 21218 USA.}
\thanks{S. S. Chandra is with the Department of Electrical and Computer Engineering, University of Texas at Dallas, Richardson, TX 75080 USA, and also with the Department of Electrical and Computer Engineering, Johns Hopkins University, Baltimore, MD 21218 USA.}
\thanks{A. Salman is with ARRAY Innovation, Bahrain.} 
\thanks{I. R. Ulgen is with the Department of Electrical and Computer Engineering, Johns Hopkins University, Baltimore, MD 21218 USA.}
\thanks{A. Mahapatra is with the Department of Electrical and Computer Engineering, Johns Hopkins University, Baltimore, MD 21218 USA.}
\thanks{C. Busso is with the Language Technologies Institute, Carnegie Mellon University, Pittsburgh PA-15213 USA (email: busso@cmu.edu).}
\thanks{B. Sisman is with the Department of Electrical and Computer Engineering, Johns Hopkins University, Baltimore, MD 21218 USA (e-mail: sisman@jhu.edu).}
\thanks{NaturalVoices:  \protect\url{https://github.com/Lab-MSP/NaturalVoices}}
\thanks{VC subsets of NaturalVoices: \protect\url{https://huggingface.co/JHU-SmileLab}}}
\maketitle
\begin{abstract}
Everyday speech conveys far more than words, it reflects who we are, how we feel, and the circumstances surrounding our interactions. Yet, most existing speech datasets are acted, limited in scale, and fail to capture the expressive richness of real-life communication. With the rise of large neural networks, several large-scale speech corpora have emerged and been widely adopted across various speech processing tasks. However, the field of voice conversion (VC) still lacks large-scale, expressive, and real-life speech resources suitable for modeling natural prosody and emotion. To fill this gap, we release NaturalVoices (NV), the first large-scale spontaneous podcast dataset specifically designed for emotion-aware voice conversion. It comprises 5,049 hours of spontaneous podcast recordings with automatic annotations for emotion (categorical and attribute-based), speech quality, transcripts, speaker identity, and sound events. The dataset captures expressive emotional variation across thousands of speakers, diverse topics, and natural speaking styles. We also provide an open-source pipeline with modular annotation tools and flexible filtering, enabling researchers to construct customized subsets for a wide range of VC tasks. Experiments demonstrate that NaturalVoices supports the development of robust and generalizable VC models capable of producing natural, expressive speech, while revealing limitations of current architectures when applied to large-scale spontaneous data. These results suggest that NaturalVoices is both a valuable resource and a challenging benchmark for advancing the field of voice conversion.
\end{abstract}

\section{Introduction}
Human speech is inherently expressive, rich in emotional nuance, and shaped by spontaneous variation in style, prosody, and interaction~\cite{scherer2003vocal, mobbs2025emotion}. Everyday communication blends linguistic content with speaker traits, affect, and social context. Yet, the datasets that dominate voice conversion (VC) and emotional VC (EVC) research rarely reflect this reality. For example, widely used datasets for emotional voice conversion consist of acted emotional speech, recorded in controlled studio environments, where conditions are very clean \cite{zhou2022emotional}. As a result, voice conversion models have been built and benchmarked on simplified data, limiting their ability to capture the richness and variability of real-world communication.  

VC aims to convert a source speaker’s voice to that of a target speaker while preserving linguistic content~\cite{qian2019autovc,9262021}, with applications in dubbing, dialogue systems, real-time voice cloning, voice assistants, and conversational agents. Early methods~\cite{toda2007voice, watts2009synthesis} relied on mapping functions trained on parallel utterances, which were costly to collect~\cite{Wrench1999mocha, toda16_interspeech}.

Deep learning has enabled powerful non-parallel approaches~\cite{xie2016kl, hsu2017voice, kaneko2018cyclegan, kameoka2018stargan, sun2016phonetic}, eliminating the need for parallel utterances across speakers or emotions with identical linguistic content. However, these models were almost exclusively trained on studio-quality data such as VCTK~\cite{yamagishi2019cstr} or CMU-Arctic~\cite{kominek04b_ssw}. Despite rapid advances in architecture design, they often show limited expressiveness, in part because the underlying training data lacks natural diversity, spontaneity, and emotion \cite{ramanujan2023connection}.  

The emergence of large-scale self-supervised~\cite{cai2025genvc,huang2021any, polyak21_interspeech,rasim_slt}, diffusion~\cite{choi2024dddm, popov2021diffusion}, codec~\cite{strecha2005codec,10446863}, and flow-matching models~\cite{yao2025stablevc,10890303}, combined with corpora like LibriTTS~\cite{zen2019libritts} and LibriSpeech~\cite{panayotov2015librispeech}, has improved intelligibility and speaker similarity. However, these datasets are largely scripted, neutral in tone, and limited in emotional coverage~\cite{Kang2023LibriheavyA5}. Consequently, even the most advanced VC architectures struggle to generate spontaneous, emotionally nuanced speech, highlighting a structural gap between the data used for training and the expressive variability of real-world speech.  

To address the limitations of neutral VC datasets, researchers have turned to emotional speech, which provides expressive variation absent from neutral recordings \cite{aihara2012gmm, xue2018voice}. This shift has sparked interest in two challenging tasks: expressive VC, which changes speaker identity while preserving emotional states \cite{du2021expressive, 10389651, gan2022iqdubbing}, and emotional VC, which converts emotional states while maintaining speaker identity \cite{5289985, zhou2021seen, 9053255}. Both tasks require accurate modeling of emotional cues and diverse speaking styles. However, most existing models \cite{du22c_interspeech, zhou2020converting, kim2020emotional} rely on the Emotional Speech Dataset (ESD) \cite{zhou2022emotional}, which contains only 30 hours of acted speech. ESD features predefined, exaggerated emotions, and lacks the subtlety and variability found in natural expression. Its limited lexical and speaker diversity further restricts emotional coverage. As a result, VC models trained on ESD often reproduce acted styles but struggle with the flexible, context-dependent emotions characteristic of real-life speech. The broader field of VC, therefore, remains constrained by its dependence on acted and controlled emotional data.  

This paper presents NaturalVoices, the first large-scale spontenous podcast dataset designed for expressive and emotional voice conversion. NaturalVoices comprises 5,049 hours of spontaneous podcast recordings spanning thousands of speakers and diverse conversational settings. Unlike acted datasets, it captures natural emotional dynamics such as shifts in anger, excitement, or sadness, as well as prosodic cues such as pitch variation, pauses, breath sounds, and voice quality changes. To make this data suitable for VC, we developed an automated processing pipeline that provides multi-level automatically generated annotations, including transcripts, speaker attributes, categorical and dimensional emotion labels, speech quality metrics, and sound events. This modular pipeline supports flexible filtering, enabling researchers to construct task-specific subsets for voice conversion.

\begin{figure}[!t]
\centering
\includegraphics[,clip,width=0.35\textwidth]{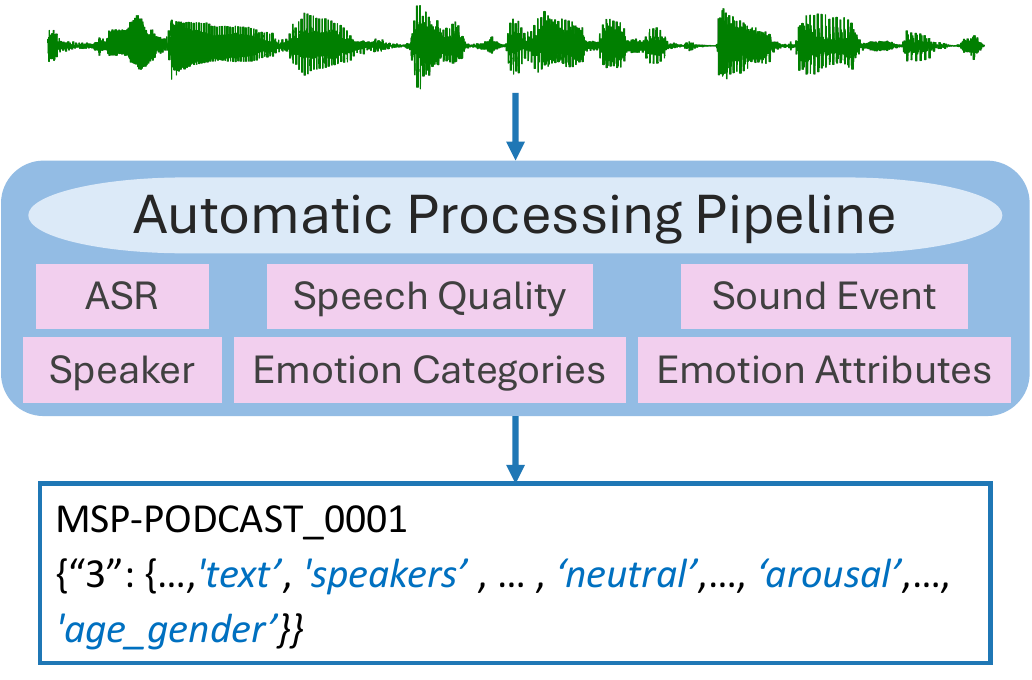}
\caption{An illustration of the proposed NaturalVoices Dataset with the automatic processing pipeline.}
\label{fig:pipeline}
\vspace{-4mm}
\end{figure}

While podcasts provide a rich source of spontaneous speech, their raw form is far from ready for VC research. Long-form episodes often include multiple speakers, background noise, and inconsistent recording quality, making them unsuitable for direct use in most VC tasks. Different applications impose different requirements: some demand clean, high-fidelity speech, while others, such as noisy-to-noisy VC \cite{10256118}, rely on realistic background conditions. Emotion-related VC further requires reliable speaker and emotion annotations. Addressing these challenges required more than simply collecting data. We built an automated processing pipeline tailored for VC (Figure~\ref{fig:pipeline}). This pipeline integrates pre-trained models and evaluation metrics to generate consistent annotations for speaker attributes (e.g., gender, identity), emotions, transcripts, speech quality, and sound events. All annotations and tools are open-sourced. Compared to our earlier work \cite{salman2024towards}, which was smaller in scale and limited to experiments with neutral data, the present dataset represents a substantial expansion in both scale and coverage, with particular emphasis on emotion-related applications.

We note that NaturalVoices is built on the same underlying podcast recordings as the MSP-Podcast corpus \cite{Busso_2025}, a widely used dataset for speech emotion recognition (SER). While the MSP-Podcast corpus includes only a manually annotated subset for speech emotion recognition (409 hours), NaturalVoices extends coverage to all speaking turns across thousands of podcast episodes for voice conversion and emotional voice conversion tasks (5,049 hours). 
We analyze NaturalVoices across multiple dimensions, including utterance duration, speaker demographics, emotion distributions, and speech quality. We then evaluate its usability for both voice conversion and emotional voice conversion by training state-of-the-art models. Results show that NaturalVoices enables the generation of natural, emotionally expressive speech with cross-domain generalization, confirming its value as a resource for advancing VC toward real-world, spontaneous speech. At the same time, the experiments reveal that current architectures are not yet able to fully leverage the dataset’s scale and variability, highlighting the need for more robust and expressive models. We believe that NaturalVoices will advance the voice conversion field by enabling models that better capture the emotion and spontaneity of real-world speech.

Our main contributions are as follows: 
\begin{itemize}
\item We present NaturalVoices, a 5,049-hour podcast dataset of real-life expressive speech collected from thousands of speakers across diverse topics. Unlike acted corpora, it captures rich emotional and stylistic variation essential for voice conversion.
\item  We release an open-source pipeline for automatic segmentation, annotation, and flexible filtering, enabling scalable and customizable data usage.  
\item We provide a comprehensive analysis of key dataset properties including speech duration, speaker diversity, emotion distributions, and quality.
\item We conduct extensive VC and EVC experiments, including out-of-domain evaluation, demonstrating that NaturalVoices not only supports effective training but also serves as a benchmark that exposes the limitations of current models.  
\item We highlight broader implications and future applications, including conversational speech synthesis, affective computing, deepfake detection, and speech enhancement.  
\end{itemize}
This paper is organized as follows: Section II reviews related work on emotional speech datasets and highlights the need for large-scale expressive resources in voice conversion. Section III describes the NaturalVoices automatic data-sourcing pipeline. Section IV provides a detailed analysis of the dataset across multiple dimensions. Section V presents voice conversion experiments. Section VI discusses broader implications and future applications. Finally, Section VII concludes the paper.
\begin{table*}[h]
\centering
\caption{Comparison of Open-Source Datasets for voice conversion containing large english subsets. Language abbreviations: En—English, Zh—Chinese, De—German, Fr—French, Ja—Japanese, Ko—Korean.}
\renewcommand{\arraystretch}{1.2} 
\resizebox{\textwidth}{!}{
\begin{tabular}{c|cccccccccc}
\hline
\textbf{Dataset} & \textbf{\makecell{Total Duration \\(Hour)}} & \textbf{\makecell{Number of \\ Speakers}} & \textbf{Type} & \textbf{\makecell{Open-source \\Pipeline}}  & \textbf{Sampling Rate} & \textbf{Segment-level Annotations} & \textbf{\makecell{Recording \\Environments}} & \textbf{\makecell{Emotion \\Labels}} & \textbf{Language(s)} \\ \hline 
VCTK\cite{yamagishi2019cstr}           &   44              & 110         & Read     & No          & 48k              & No  & Studio & No & En\\ 
CMU-Arctic\cite{kominek04b_ssw}       & 7.18                &   7       &Read      & No         & 16k                 &No &Studio  &No & En\\
VCC 2016\cite{toda16_interspeech}            &  2.11               & 10        & Read     & No          & 16k              & No  & Studio & No & En\\
VCC 2018\cite{lorenzo-trueba2018voice}            & 1.35                & 12         & Read     & No          & 16k              & No  & Studio & No & En\\
VCC 2020\cite{Yi2020}            & 1.41                   & 8         & Read     & No          & 16k              & No  & Studio & No & En\\
Libritts\cite{zen2019libritts}       &  585               &  2,456        &Read       & No          & 24k                  &No &Studio  & No& En\\

Emilia \cite{he2024emilia}     &     101K            & No spk labels/count          &        Spontaneous&    Yes   &          24k      & No& In-the-wild & No& En/Zh/De/Fr/Ja/Ko\\

AutoprepWild\cite{yu2024autoprep}      &           39      &  48     &Spontaneous       &  No          &  24k/44.1k                  & Yes   &In-the-wild  &No& En\\
LibriLight\cite{kahn2020libri}
& 60k & 7439 & Read & No & 16k & No & Studio & No & En
\\
GigaSpeech \cite{chen2021gigaspeech} & 10k & No spk label/count& Read/Spontaneous & No & 16k& Limited & Studio/In-the-wild & No &En \\
\hline
ESD\cite{zhou2022emotional}          & $15^*$                &  10        & Read     &No           & 16k             &No &Studio  & Yes& En/Zh\\ 
Expresso\cite{nguyen2023expresso}     & 40                & 4         &Read and Improvised      &No           &48k             &No & Studio &Yes & En\\ 
MSP-Podcast 2.0\cite{8003425}   &    409             &3,641         &Spontaneous      & No          &16k            &Yes &In-the-wild &Yes & En\\ \hline
NaturalVoices-v0\cite{salman2024towards}  &3,846                 &$>$2,467          &Spontaneous      &Yes           & 16k              &Yes &In-the-wild &Yes & En\\ 
\textbf{NaturalVoices (proposed)} & 5,049               &$>$ 2,670        &Spontaneous      &Yes           & 16k, 44.1k and etc              &Yes &In-the-wild &Yes& En \\ 
\hline
\end{tabular}}
\label{tab: comparsion}
\vspace{-5mm}
\end{table*}
\vspace{-5mm}
\section{Related Work}
\subsection{Limitations of Emotional Speech Datasets for VC}
A central obstacle for emotional VC is the lack of suitable datasets. A fundamental requirement is speaker diversity, which allows models to learn varied speaker characteristics, perform accurate conversion across identities, and generalize to unseen speakers~\cite{zhou2022emotional}. Another major challenge is capturing the complexity of emotional expression~\cite{chandra2024exploring}. The same emotion can be expressed differently across individuals~\cite{scherer2003vocal}, meanings shift with context, and emotions vary in intensity or blend into mixed states~\cite{plutchik2013theories}. Dimensional representations such as arousal, valence, and dominance~\cite{barrett2006solving} offer a richer alternative, but are difficult to annotate reliably at scale. Together, these factors make it especially difficult to construct corpora that fully capture the richness of emotional speech.  

Collecting such corpora in controlled environments is also costly. As a result, researchers have explored in-the-wild sources such as podcasts~\cite{8003425} and YouTube~\cite{li2023yodas}. However, these strategies introduce new difficulties: utterance lengths are highly variable, background conditions are inconsistent, and large-scale annotation is expensive. Overlapping or recurring speakers complicate speaker identification, while reliable emotion labeling often requires costly human annotation or robust automated tools. Many emotional speech datasets originally developed for speech emotion recognition (SER) lack the scale, linguistic coverage, and recording quality required for VC. Most SER corpora are relatively small~\cite{james18_interspeech,livingstone2018ryerson,cao2014crema}, and some corpora, such as IEMOCAP~\cite{Busso2008} and CREMA-D~\cite{cao2014crema}, have a very limited number of speakers, which reduces their usability for voice conversion.

Because of these challenges, voice conversion research has been forced to rely on acted, studio-quality datasets in which emotions are simulated and conditions are artificially clean. This dependence has shaped the entire field: models are designed and benchmarked on simplified data, and as a result, they struggle to capture the spontaneity and expressiveness of real-world emotional speech.  
\vspace{-3mm}
\subsection{Datasets for Voice Conversion}
Table~\ref{tab: comparsion} summarizes widely used open-source English datasets for VC, highlighting their scale, speaker coverage, and annotations. Most existing resources fall into two categories: neutral speech datasets and emotional speech datasets. Neutral datasets, such as VCTK~\cite{yamagishi2019cstr}, CMU-Arctic~\cite{kominek04b_ssw}, and the Voice Conversion Challenge (VCC) series~\cite{toda16_interspeech, lorenzo-trueba2018voice, Yi2020}, provide high-quality studio recordings of read or scripted speech. While widely used and effective for benchmarking, these corpora lack the spontaneity, diversity, and emotional richness of real-world communication. Larger text-to-speech corpora such as LibriTTS (585 hours, 2,456 speakers)~\cite{zen2019libritts} and LibriLight (60k hours, 7439 speakers) \cite{kahn2020libri} have broadened speaker coverage, but they remain scripted and neutral in style. To enable expressive VC, several emotional datasets have been introduced. The Emotional Speech Dataset (ESD)~\cite{zhou2022emotional} (15 hours, 10 speakers) is among the most widely used, but its acted emotions limit spontaneity and subtlety. Expresso~\cite{nguyen2023expresso} (40 hours, 4 speakers) includes both read and improvised speech, but the small number of speakers restricts generalization. More recently, large-scale in-the-wild datasets have been constructed for speech generation tasks. In our study, we found that these datasets either lack detailed segment-level annotations \cite{he2024emilia,chen2021gigaspeech} or do not provide the flexibility for data filtering and subset selection through an open-source pipeline \cite{yu2024autoprep}. The MSP-Podcast corpus~\cite{8003425}, originally designed for SER, also provides large-scale in-the-wild data but lacks speech quality assessments critical for VC.  

Together, these corpora have shaped the field by providing clean, controlled benchmarks, but they are fundamentally mismatched with the demands of real-world expressive speech. Models trained on these resources learn to reproduce acted or scripted conditions, but they struggle with the variability, spontaneity, and emotional dynamics of natural communication. This underscores the need for large-scale, naturalistic resources such as NaturalVoices.

\begin{figure*}[!ht]
\centering
\includegraphics[,clip,width=0.9\textwidth]{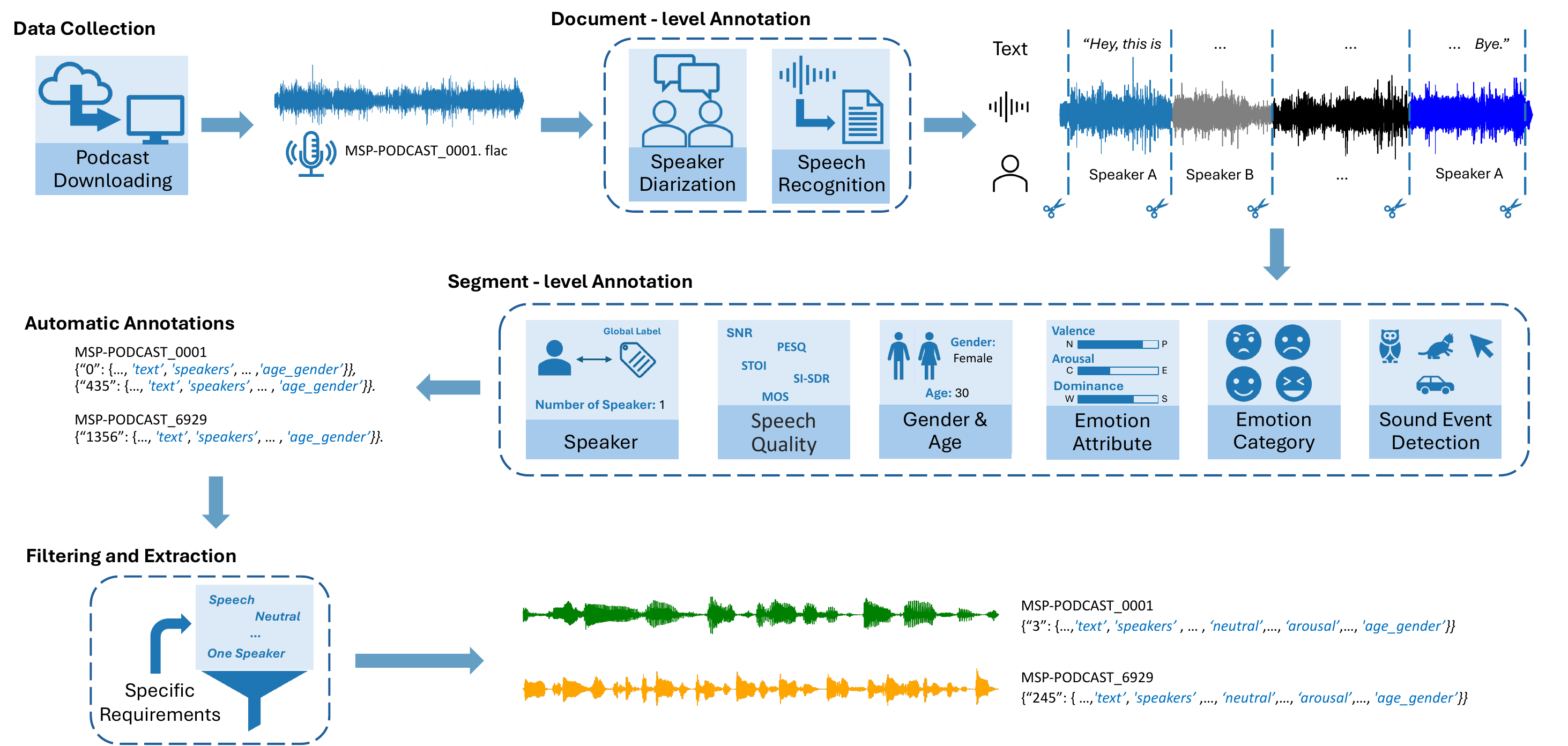}
\caption{An illustration of our pipeline processing NaturalVoices dataset with various modules, which includes speaker diarization, speech recognition, speech quality evaluation, emotion attribute and category prediction, and sound event detection.}
\label{fig:pipelinestruct}
\vspace{-5mm}
\end{figure*} 
\vspace{-3mm}
\subsection{Large Models for Expressive Speech Synthesis and Conversion}
Recent advances in generative modeling have driven major advances in speech synthesis and VC. In NLP, large language models such as GPT~\cite{openai2023gpt4} and BERT~\cite{devlin2018bert} produce coherent, context-rich text, while in vision, diffusion models such as Stable Diffusion~\cite{rombach2022high} and DALL-E~\cite{ramesh2021dalle} generate realistic images. Similar trends are observed in TTS: transformer-based architectures~\cite{li2019neural}, LLM-based systems~\cite{wang2023neural,zhang2023speechlm,casanova2022yourtts,shi2022naturalspeech}, and diffusion models~\cite{ijcai2022p577,kong2021diffwave} have significantly improved speech quality, naturalness, and expressive control. Emotional TTS systems now leverage large models for more expressive synthesis~\cite{osman2022emo,yoon22b_interspeech,wu2024laugh,du2024cosyvoice}. However, their effectiveness is constrained by the lack of large, diverse emotional datasets. Inspired by TTS, similar architectures have been adopted in emotional VC. For example, Durflex-EVC~\cite{oh2024durflex} employs transformer-based style encoders, while DEVC~\cite{du2024converting} integrates self-supervised features with a diffusion-based decoder for expressive synthesis. These advances highlight the importance of large models and scaling the data to meet the modeling needs. 
\vspace{-3mm}
\subsection{Summary of Research Gap}
We identify key limitations in existing emotional speech resources for voice conversion.

\begin{itemize}
\item 
Most available VC datasets are small and recorded in controlled environments. Few provide detailed annotations such as speaker traits, emotion labels, prosody, and acoustic conditions (e.g., SNR, background noise).

\item 
Current VC models often fail to generalize to real-world speech, and there is a lack of benchmark data to evaluate robustness under such conditions.

\item 
Due to the lack of large-scale emotional speech, most VC research still relies on studio-quality neutral recordings. This dependence on acted and controlled data has constrained the field, preventing progress on more realistic scenarios that require robust modeling of spontaneous emotional speech.
\item 
Most existing datasets are generally pre-processed before release, with noise and variability removed. While this simplifies use, it restricts flexibility for downstream research, as users cannot easily select or customize subsets best suited to their tasks. 
\end{itemize}
To address these gaps, we present NaturalVoices dataset, which provides large-scale, real-life emotional speech with comprehensive annotations. We also present an automatic pipeline that enables flexible filtering and customization, allowing researchers to build task-specific subsets for diverse VC scenarios.
\vspace{-3mm}
\section{Automatic Data-sourcing Pipeline in NaturalVoices Dataset}
NaturalVoices consists of two main components: the podcast audio itself and an automated data-sourcing pipeline that enriches each episode with detailed metadata while preparing the data for speech-related tasks, particularly VC. As illustrated in Figure~\ref{fig:pipelinestruct}, the pipeline has four main stages:

\begin{enumerate}
\item Data Collection
\item Document-level Annotation 
\item Segment-level Annotation 
\item Filtering and Extraction
\end{enumerate}

This section provides a detailed overview of each stage and its corresponding modules, highlighting their roles in processing podcast episodes for downstream applications. The novelty of our pipeline lies in its structured design for automatically annotating large-scale, real-life emotional speech with rich, multi-level metadata. Real-world speech encodes diverse information such as linguistic content, emotional states, speaker traits, and background context that is often underutilized due to the high cost of manual labeling. Our approach leverages pre-trained models to produce diverse annotations at scale, with particular focus on emotional and stylistic cues. This flexible framework enables data preparation and supports a broad range of downstream applications, especially emotion-aware voice conversion and expressive speech generation.
\vspace{-3mm}
\subsection{Data Collection}
We collected over 6,790 podcast episodes from the internet, all available under Creative Commons licenses. To optimize storage, the recordings were converted to FLAC format, which provides more efficient compression than the WAV format used in NaturalVoices-v0~\cite{salman2024towards}. Each audio file has an average duration of approximately 45 minutes. In addition to the 16kHz downsampled files provided in the previous release \cite{salman2024towards}, NaturalVoices also includes recordings at their original sample rates. This allows for greater flexibility in research applications, particularly for those requiring high-fidelity audio.


\subsection{Document-level Annotation} 
From this point onward, we refer to each podcast episode as a document, and the individual speech utterances within an episode as segments.
\subsubsection{Automatic Speech Recognition (ASR)} We apply the Faster-Whisper model\footnote{\url{https://github.com/SYSTRAN/faster-whisper}}, an optimized implementation of Whisper \cite{radford2023robust} built with CTranslate2, to perform automatic speech recognition and segment each podcast episode into short, utterance-level clips. For each segment, the model outputs transcripts, language identifiers, and confidence scores. As illustrated in Figure~\ref{fig:example}, this step produces the fields ``Start,'' ``End,'' ``Text,'' and ``ASR\_CONF,'' which capture segment boundaries, transcribed content, and the ASR model’s confidence in the generated transcription. 

\subsubsection{Speaker Diarization} We utilize PyAnnote\footnote{\url{https://github.com/pyannote/pyannote-audio}} \cite{Bredin23, Plaquet23}, trained on large-scale speaker diarization datasets, to estimate the number of speakers and assign speaker labels within each podcast episode. As shown in Figure~\ref{fig:example}, the ``Speakers'' field specifies each speaker’s time span (e.g., 0.10--2.00) and the assigned label (e.g., ``SPEAKER\_00,'' ``SPEAKER\_01''). These labels distinguish speakers within a single episode but do not resolve speaker identity across episodes.  As shown in Figure~\ref{fig:pipelinestruct}, this process generates a text transcription for each segment, along with local speaker labels and precise time boundaries within each podcast document.
\vspace{-4mm}
\subsection{Segment-level Annotation}

The segment-level annotation stage builds on the initial document-level annotation by adding detailed information about the acoustic and linguistic properties of each segment. The main components involved in this process are listed below. 

\subsubsection{Speaker} We incorporate global speaker identities from the MSP-Podcast corpus \cite{8003425}, which provides human-annotated utterance-level timestamps. Since our dataset includes time-stamped segments with local speaker labels from diarization, we perform a two-stage mapping process:  

\begin{itemize}
    \item Mapping: Each segment is aligned with MSP-Podcast annotations by verifying that it originates from the same podcast and falls within the labeled time boundaries. Matching segments are assigned the corresponding global speaker label.  
    \item Mapping+Prediction: Global speaker labels are linked to local diarization labels. If a global speaker uniquely corresponds to a local diarization label within an episode, the global label is propagated to all matching segments.  
\end{itemize}  

As shown in Figure~\ref{fig:example}, the ``Global Speaker'' field records both the assigned label (e.g., ``30'') and the annotation method (``Mapping'' or ``Mapping+Prediction''). To maintain reliability, no global label is assigned when inconsistencies occur (e.g., one global speaker mapping to multiple diarization labels). By combining human-annotated speaker identities with automatic diarization, our approach ensures consistent speaker labeling across episodes. This unified annotation is essential for tasks such as voice conversion, where stable speaker identity is critical for training and evaluation.

\subsubsection{Speech Quality}
\begin{figure}[t]
\centering
\includegraphics[,clip,width=0.4\textwidth]{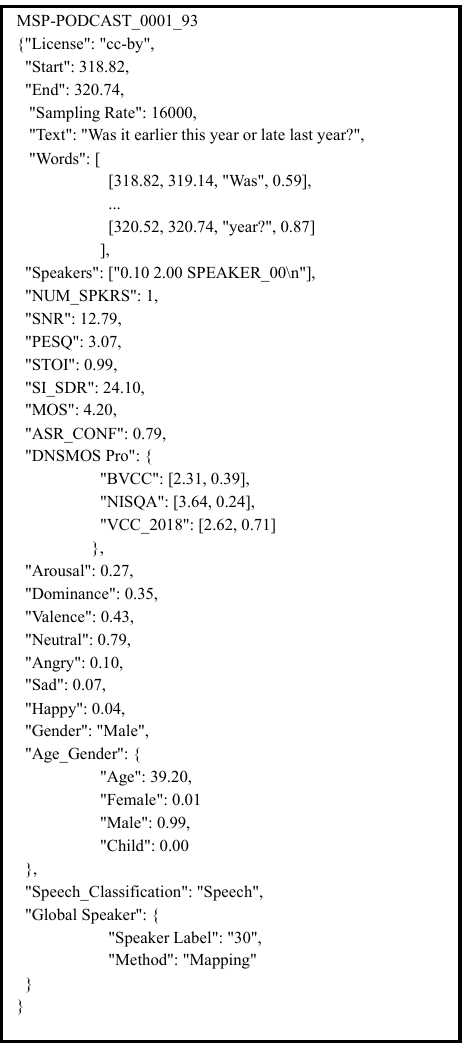}
\caption{Segment-level annotation from the NaturalVoices dataset. Shown is a 2 second speech segment from document MSP-PODCAST\_0001\_93. Each segment entry includes speech quality metrics, emotion and speaker attributes, and additional metadata describing the acoustic and contextual characteristics of the audio.}
\label{fig:example}
\vspace{-5mm}
\end{figure}

Because podcast audio is recorded in real-world conditions, it naturally exhibits variability in quality. To characterize this variability, we design a multi-metric module that evaluates three key dimensions: \textit{perceived quality}, \textit{intelligibility}, and \textit{noise level}. We leverage Torchaudio-Squim\cite{kumar2023torchaudio}\footnote{\url{https://pytorch.org/audio/main/tutorials/squim_tutorial.html}}, which provides interfaces and pre-trained models for several standard measures:  
\begin{itemize}  
    \item PESQ \cite{rix2001perceptual}: Perceptual evaluation of speech quality. 
    \item STOI \cite{5495701}: Short-time objective intelligibility.  
    \item SI-SDR \cite{le2019sdr}: Scale-invariant signal-to-distortion ratio.  
    \item MOS \cite{manocha22c_interspeech}: Mean opinion score, estimating human-perceived quality (1–5).  
\end{itemize}  
These outputs are recorded in the ``PESQ,'' ``STOI,'' ``SI\_SDR,'' and ``MOS'' fields, as illustrated in Figure~\ref{fig:example}.  

To assess background noise, we compute signal-to-noise ratio (SNR) \cite{kim08e_interspeech} using the WADA-SNR method\footnote{\url{https://gist.github.com/johnmeade/d8d2c67b87cda95cd253f55c21387e75}}. In addition, we apply DNSMOS Pro\footnote{\url{https://github.com/fcumlin/DNSMOSPro}} \cite{cumlin24_interspeech}, a neural metric that predicts noise suppression quality based on DNSMOS \cite{9414878}. As shown in the ``DNSMOS Pro'' field of Figure~\ref{fig:example}, the model outputs a mean score and variance, trained on BVCC \cite{huang22f_interspeech}, NISQA \cite{mittag21_interspeech}, and VCC 2018 \cite{lorenzo-trueba2018voice}. Together, these measures provide a rich, multi-dimensional characterization of speech quality, ensuring that segments can be flexibly selected for tasks requiring specific quality conditions.  

\subsubsection{Gender and Age}  
We use a pre-trained model\footnote{\url{https://github.com/audeering/w2v2-age-gender-how-to}} \cite{burkhardt2023speech}, trained on demographic-labeled corpora, to predict speaker gender and estimate speaker age. 

\subsubsection{Emotion Categories} Categorical emotions are predicted using the PEFT-SER model\footnote{\url{https://github.com/usc-sail/peft-ser}} \cite{feng2023peft}. This model is based on WavLM with LoRA fine-tuning, and classifies speech into four categories: \textit{anger}, \textit{sadness}, \textit{happiness}, and \textit{neutral}. It is trained on multiple emotional corpora, including IEMOCAP \cite{Busso2008}, MSP-Improv \cite{Busso2017}, MSP-Podcast \cite{8003425}, and CREMA-D \cite{Cao2014}. These outputs provide consistent categorical emotion labels for each segment, complementing the continuous emotion attributes described in the next section.  
\subsubsection{Emotion Attributes} 
To provide richer emotional information, our pipeline also provides continuous emotional attributes that represent affective states in a multidimensional space: 1) valence (positivity or negativity of the emotional tone), 2) arousal (level of activation, ranging from calm to excited), 3) dominance (degree of control or assertiveness, from weak to strong). Unlike emotion categories, emotion attributes represent emotional information in a multidimensional continuous space \cite{Busso_2025}. We utilize a pre-trained regression-based WavLM model\footnote{https://huggingface.co/3loi/SER-Odyssey-Baseline-WavLM-Multi-Attributes}\cite{goncalves24_odyssey}, which was trained on emotion-labeled speech data, to capture the speaker's emotional state across these three dimensions.

\subsubsection{Sound Event Detection} Non-speech events such as background noise and music are identified using the pre-trained AST model\footnote{\url{https://github.com/YuanGongND/ast}} \cite{gong_psla, gong21b_interspeech}, trained on large-scale audio corpora with labeled sound events. The model supports detection of 527 sound classes (e.g., honking, alarms, animal noises), enabling segment-level annotations of background context. These labels are particularly useful for studying robustness in voice conversion under real-world acoustic conditions.  
\vspace{-2mm}
\subsection{Filtering and Extraction}
The rich annotations in NaturalVoices enable users to filter data by specific criteria and extract tailored subsets for their applications. In Section~5, we demonstrate how these annotations can be used to construct customized datasets for various VC tasks, illustrating practical examples of task-driven filtering strategies.  

\section{NaturalVoices: A Large-Scale Dataset of Spontaneous, Emotionally Rich Speech}

NaturalVoices is a large-scale, richly annotated podcast dataset comprising 5,049 hours of spontaneous, in-the-wild speech from 6,790 episodes.\footnote{ NaturalVoices is available at https://github.com/3loi/NaturalVoices} Unlike existing resources, it combines emotional expressiveness, speaker diversity, and real-world acoustic variability, making it uniquely suited for expressive voice conversion and emotional speech modeling. Building on our previous release, NaturalVoices-v0 \cite{salman2024towards}, this version expands the dataset with substantially more recordings and an improved automated pipeline, as summarized in Table~\ref{tab: comparsion}. New annotations include sampling rate, license type, speech quality metrics, and emotion-related labels, extending its use across a wide range of voice conversion and general speech processing tasks.  

In this section, we provide a comprehensive analysis of NaturalVoices across multiple dimensions, with a focus on emotional expressiveness, conversational structure, and linguistic variety. Despite variability in audio quality, these real-world characteristics are essential for developing robust, generalizable models.
\begin{figure*}[t]
    \centering
    \subfloat[Sampling Rate]{\includegraphics[width=0.44\textwidth]{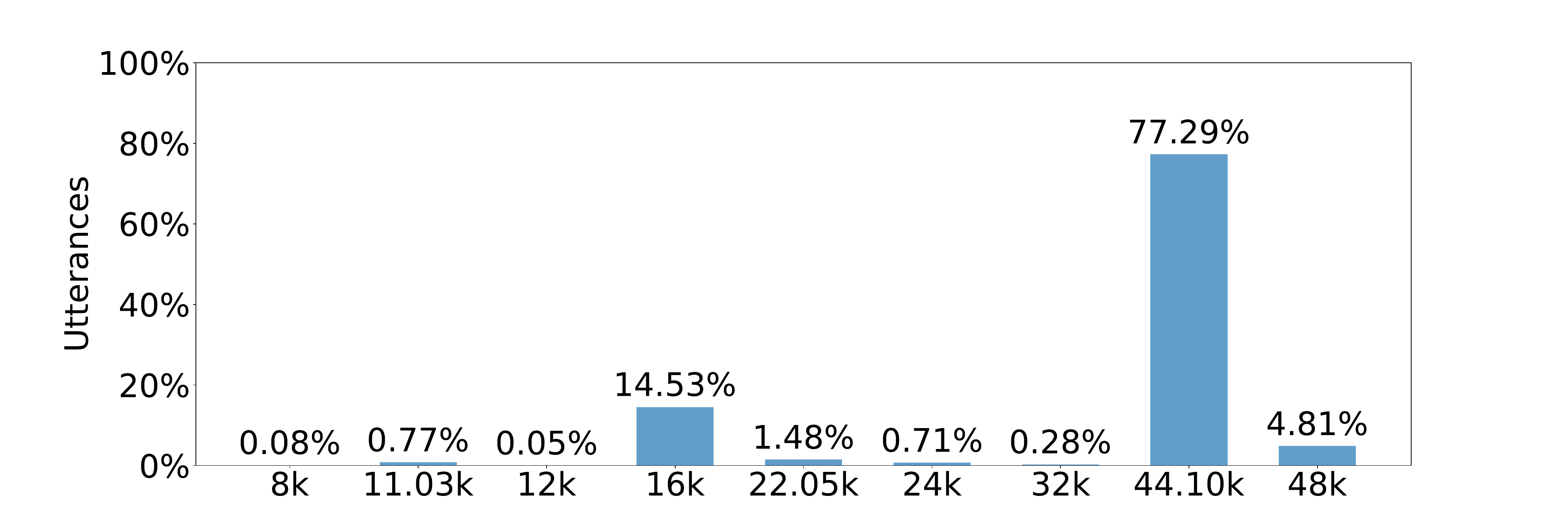}%
    } 
    \subfloat[Duration]{\includegraphics[width=0.29\textwidth]{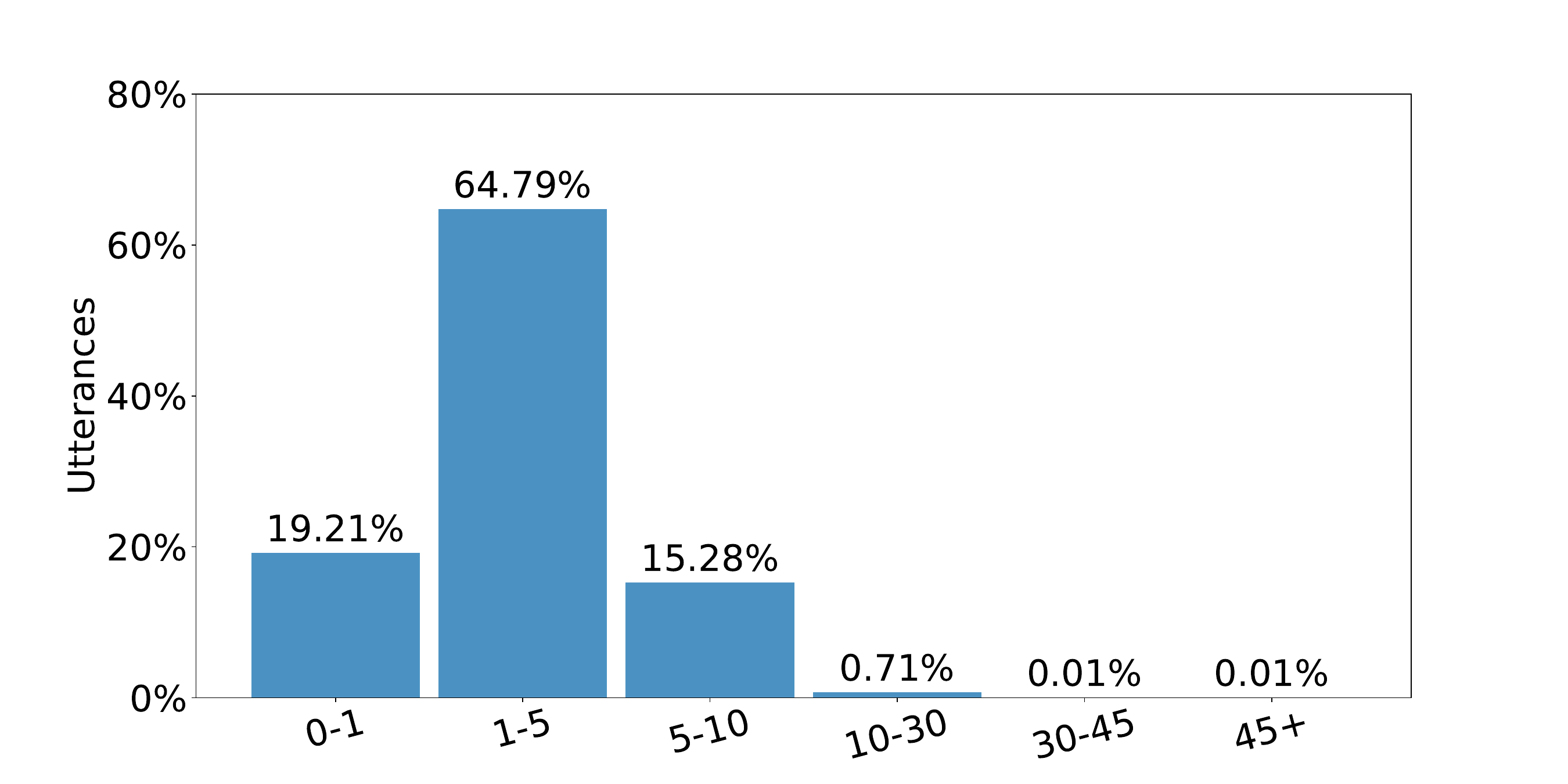}%
    } 
    \subfloat[Gender]{\includegraphics[width=0.28\textwidth]{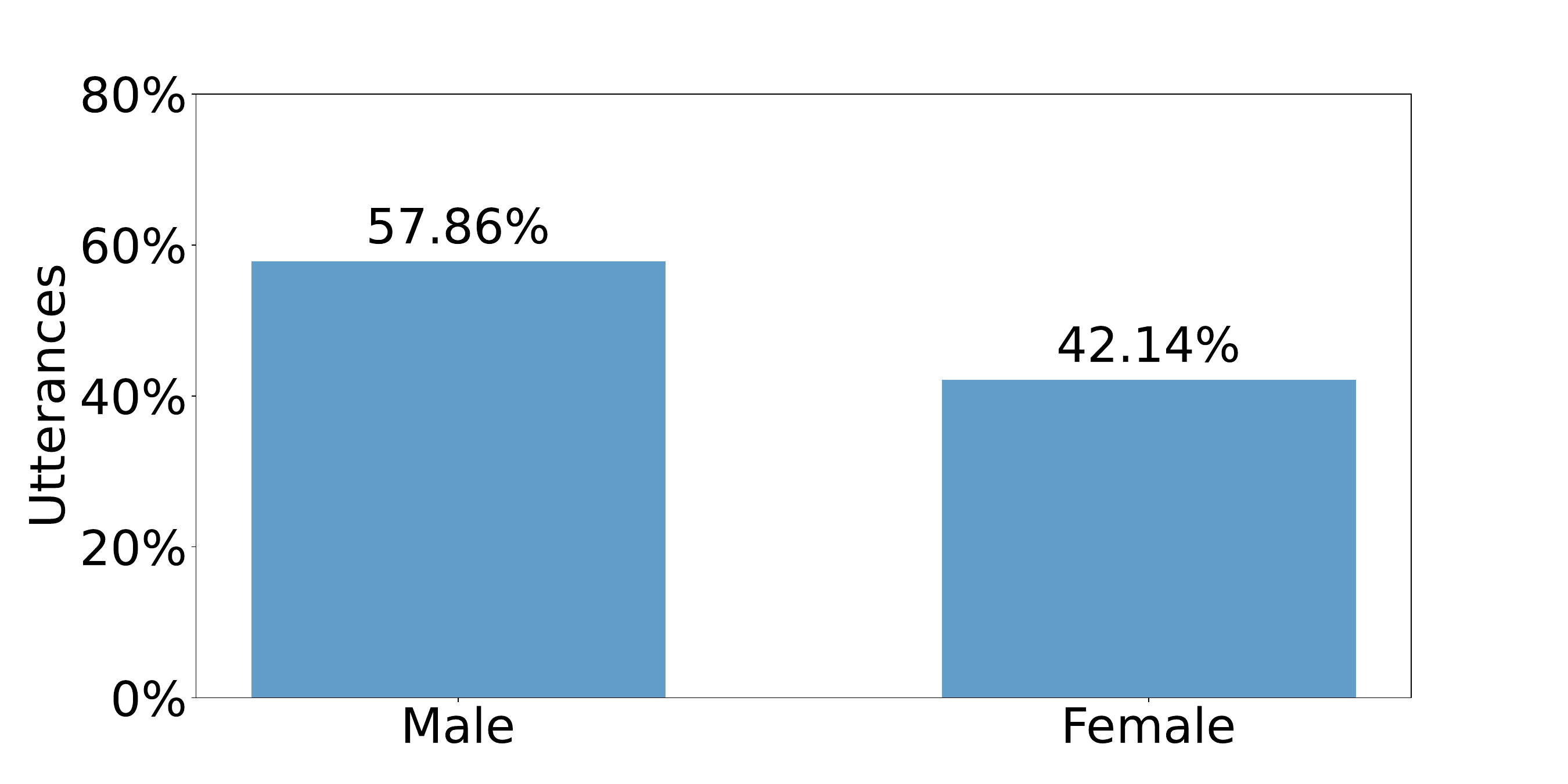}%
    }
    \caption{Distributions of sampling rates, utterance durations, and speaker gender in NaturalVoices.}
    \label{fig:gender duration}
\end{figure*}
\subsection{Why Podcast Speech for Voice Conversion}

Podcast data is a rich source of spontaneous, expressive speech data. While it has been constructed and used for ASR\cite{clifton2020spotify}, TTS \cite{szekely2019spontaneous} and SER \cite{Busso_2025}, it remains underexplored by the VC community.
To address this gap, we introduce NaturalVoices, built from podcast speech and providing a strong foundation for voice conversion research. Podcasts naturally combine spontaneity with relatively high recording quality. Unlike scripted or studio-acted speech, podcasts capture authentic, emotionally rich conversations where the speaker’s delivery is genuinely aligned with the content being expressed. The diversity of hosts and guests spanning different ages, accents, and cultural backgrounds exposes models to a wide range of vocal characteristics, which is essential for achieving true generalizability in voice conversion systems. Additionally, podcast discussions often involve deliberate reasoning, debate, or storytelling, allowing models to learn expressive yet coherent prosodic patterns.
\vspace{-3mm}
\subsection{Dataset Characteristics: Sampling Rate, Duration, Gender}
\subsubsection{Sampling Rates}
NaturalVoices includes recordings at multiple sampling rates to support a wide range of speech processing tasks. As shown in Figure~\ref{fig:gender duration}(a), most utterances (77.29\%) are recorded at 44.1 kHz, providing high-fidelity audio suitable for expressive and high-quality speech generation. The dataset also contains recordings at 16 kHz (14.53\%) and 48 kHz (4.81\%), allowing users to freely downsample for computationally efficient modeling or retain original high-resolution signals for tasks that require enhanced fidelity. The availability of diverse sampling rates increases the dataset’s flexibility for both efficient and high-fidelity applications.

\subsubsection{Utterance Durations}
Figure~\ref{fig:gender duration}(b) shows that 99.28\% of utterances range from 1–10 seconds, with about 65\% concentrated in the 1–5 second interval. This distribution aligns with other widely used VC datasets \cite{veaux2017cstr}. The inclusion of longer utterances (Table~\ref{table:speech_duration}) further supports advanced tasks such as long-form speech synthesis \cite{park2024long}.  

\begin{table}[!t]
\centering
\caption{Cumulative duration (in hours) of utterances across longer length intervals in NaturalVoices, filtered for speech labels only. This provides additional details for utterances longer than 30 seconds, complementing the duration distribution shown in Figure \ref{fig:gender duration}(a).}
\resizebox{0.33\textwidth}{!}{%
\begin{tabular}{c|c}
\hline
\textbf{Sentence Length (s)} & \textbf{Duration (h)} \\ \hline
30-60 & 6.66 \\ 
60-120 & 3.19 \\ 
More than 120 & 4.52 \\ \hline
\end{tabular}%
}
\label{table:speech_duration}
\end{table}

\subsubsection{Speaker Gender}
As shown in Figure~\ref{fig:gender duration}(c), the dataset has a balanced gender distribution: 54.73\% of utterances are from male speakers and 45.27\% from female speakers. This balance enhances the dataset’s diversity and mitigates gender-related biases, making it well suited for fair and representative speaker modeling.  

\begin{figure}[!t]
    \centering
    \subfloat[Word cloud]{\includegraphics[width=0.66\columnwidth]{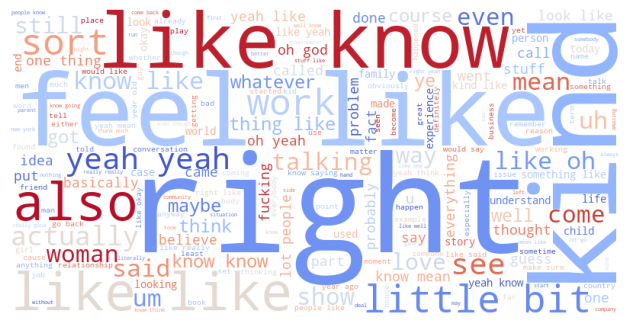}%
    }\\
    \subfloat[Text Sentiment Classification]{\includegraphics[width=0.66\columnwidth]{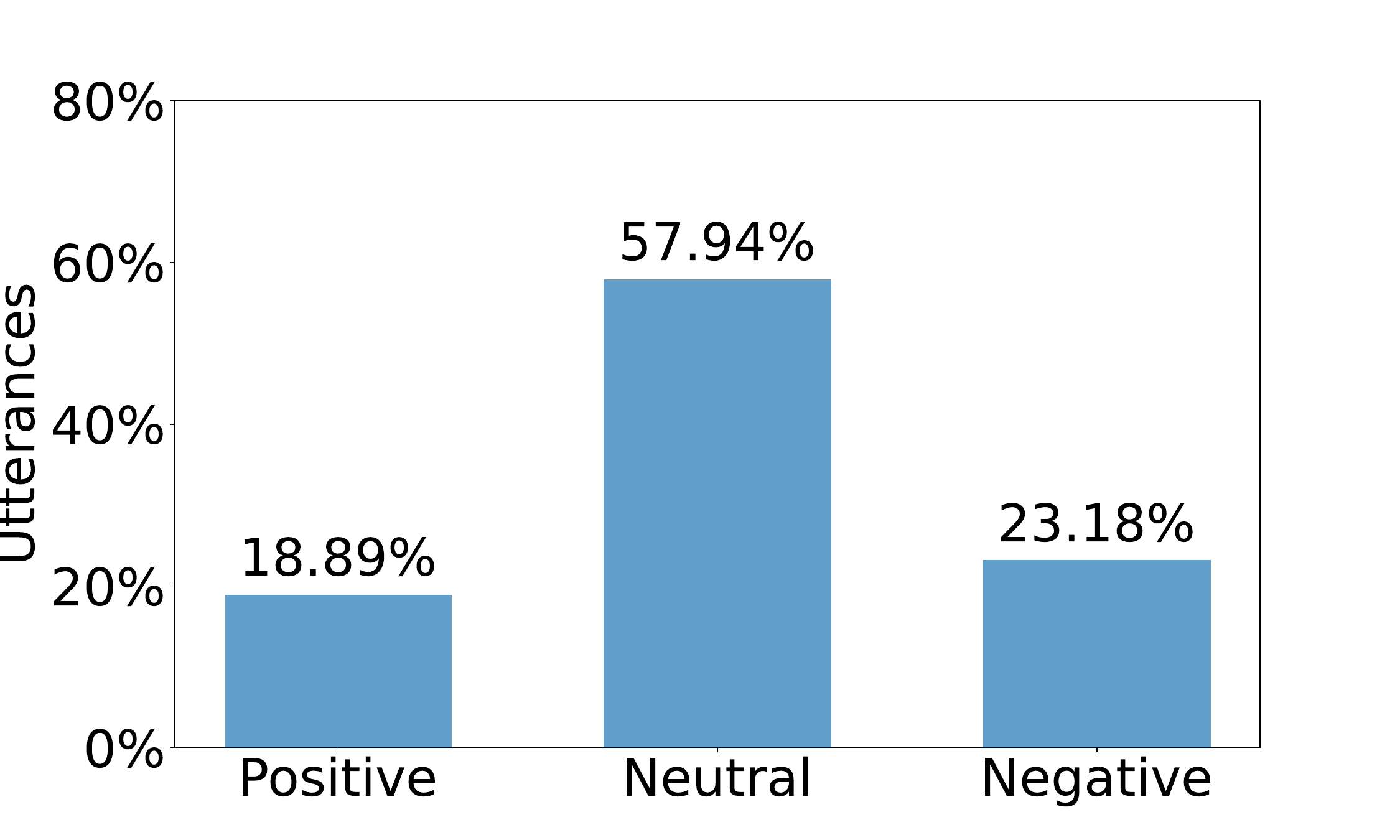}
    }\\

    \caption{Lexical and sentiment analysis of NaturalVoices: (a) word cloud of frequent terms, (b) distribution of text sentiment categories.}
\label{fig:Spoken_Content_Analysis}
\end{figure}

\begin{figure}[t]
\centering
\includegraphics[,clip,width=0.66\columnwidth]{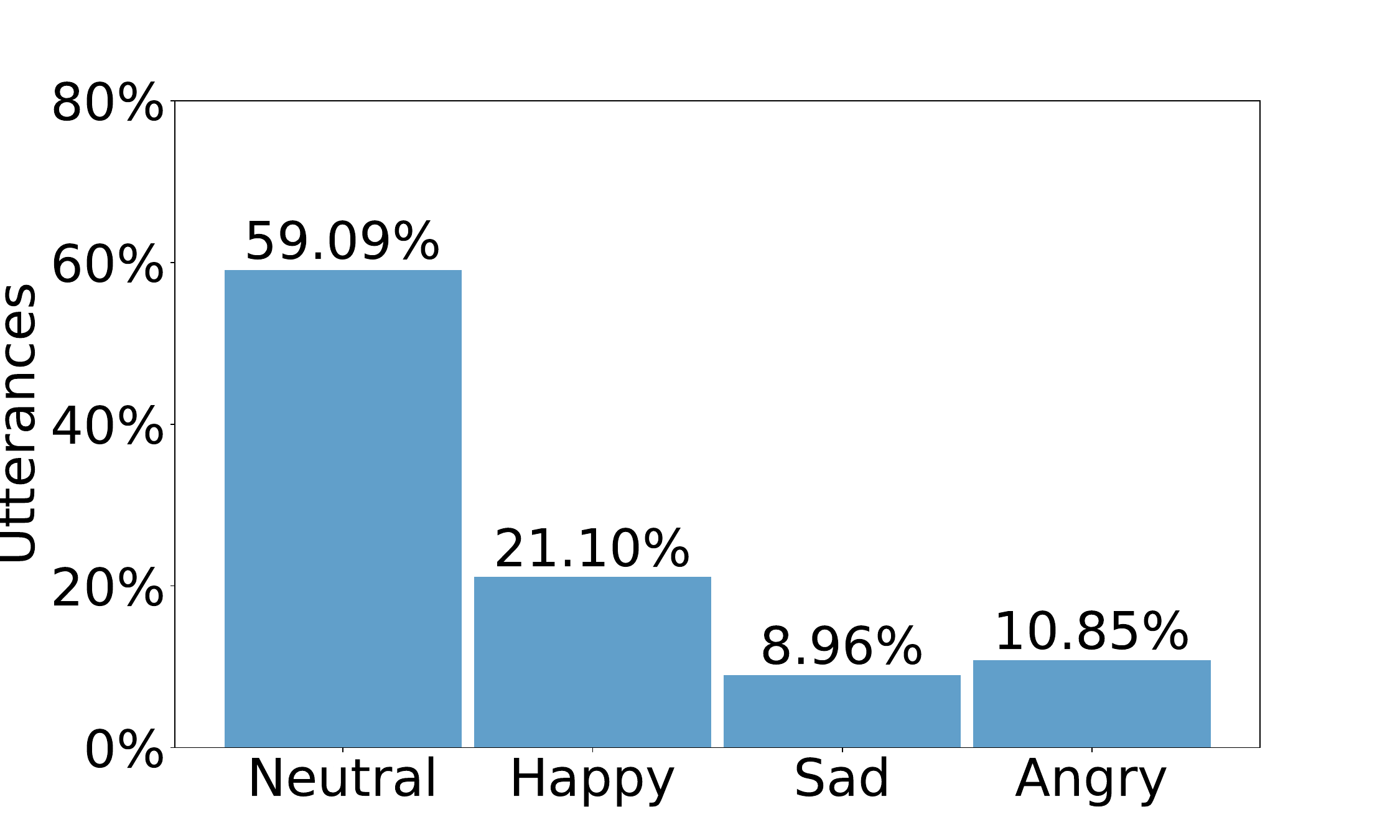}
\caption{Distribution of emotion categories in NaturalVoices.}
\label{fig:emotion_ca}
\vspace{-5mm}
\end{figure}
\vspace{-4mm}
\subsection{Spoken Content Analysis}
We analyze the lexical distribution of NaturalVoices using a word cloud, shown in Figure~\ref{fig:Spoken_Content_Analysis}(a). The most frequent words include colloquial terms such as ``like,'' ``know,'' ``right,'' and ``feel.'' Filler words (e.g., ``um'') and casual expressions (e.g., ``sort of,'' ``little bit'') further illustrate the spontaneous and conversational nature of the speech. These naturalistic features make the dataset particularly well suited for research on spontaneous and expressive speech modeling, as well as conversational analysis.  
\vspace{-4mm}
\subsection{Emotional Characteristics}
Emotion is a fundamental component of natural speech, as human conversations inherently blend emotional states with communicative intent. To provide a comprehensive view of the emotional content in NaturalVoices, we analyze emotion-related information derived from both the text and the speech audio. This analysis underscores the suitability of the data set for developing expressive speech models capable of capturing emotional nuances more effectively.

\begin{figure*}[t]
    \centering
    \subfloat[Arousal]{\includegraphics[width=0.33\textwidth]{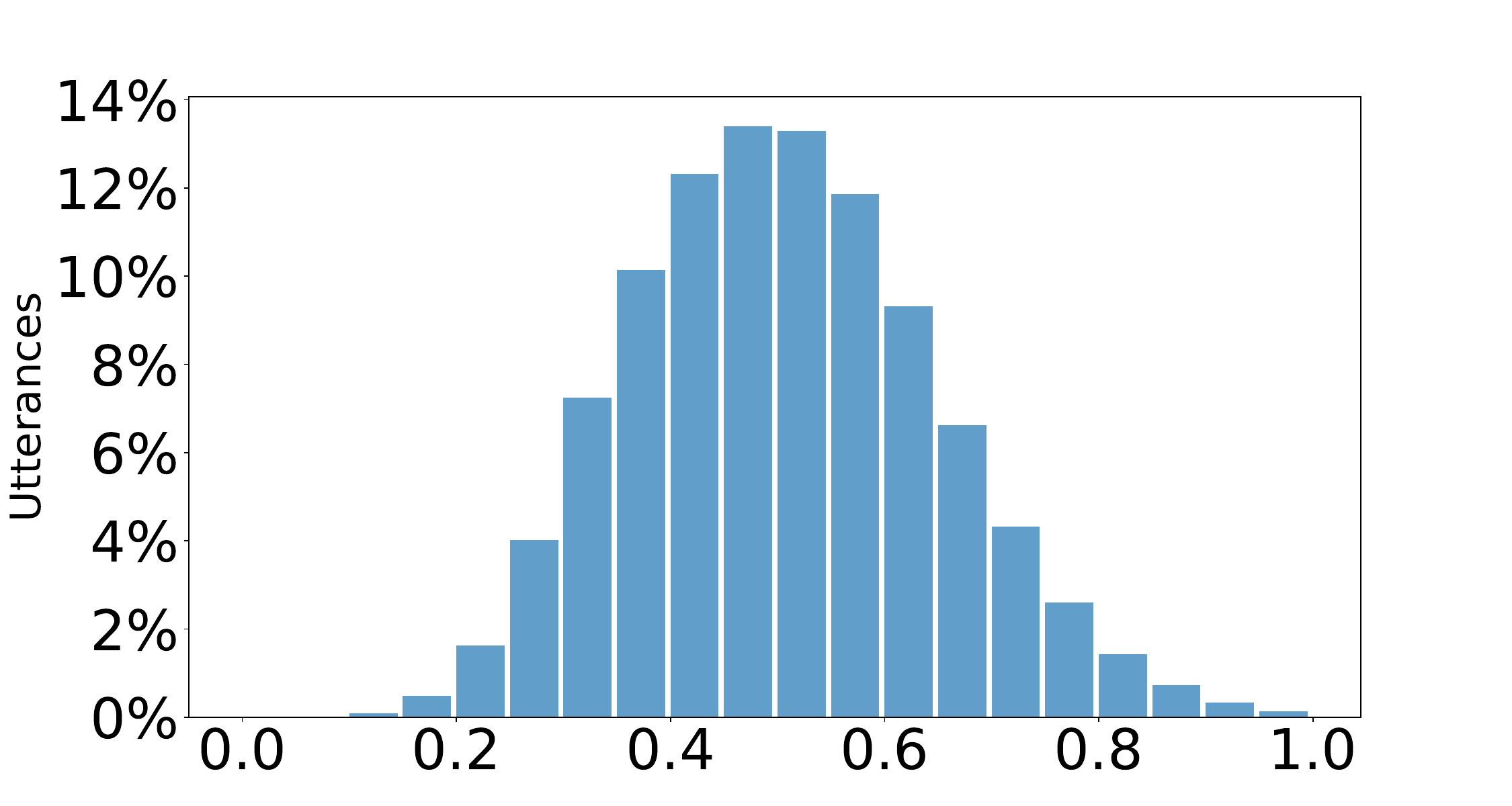}%
    }
    \subfloat[Dominance]{\includegraphics[width=0.33\textwidth]{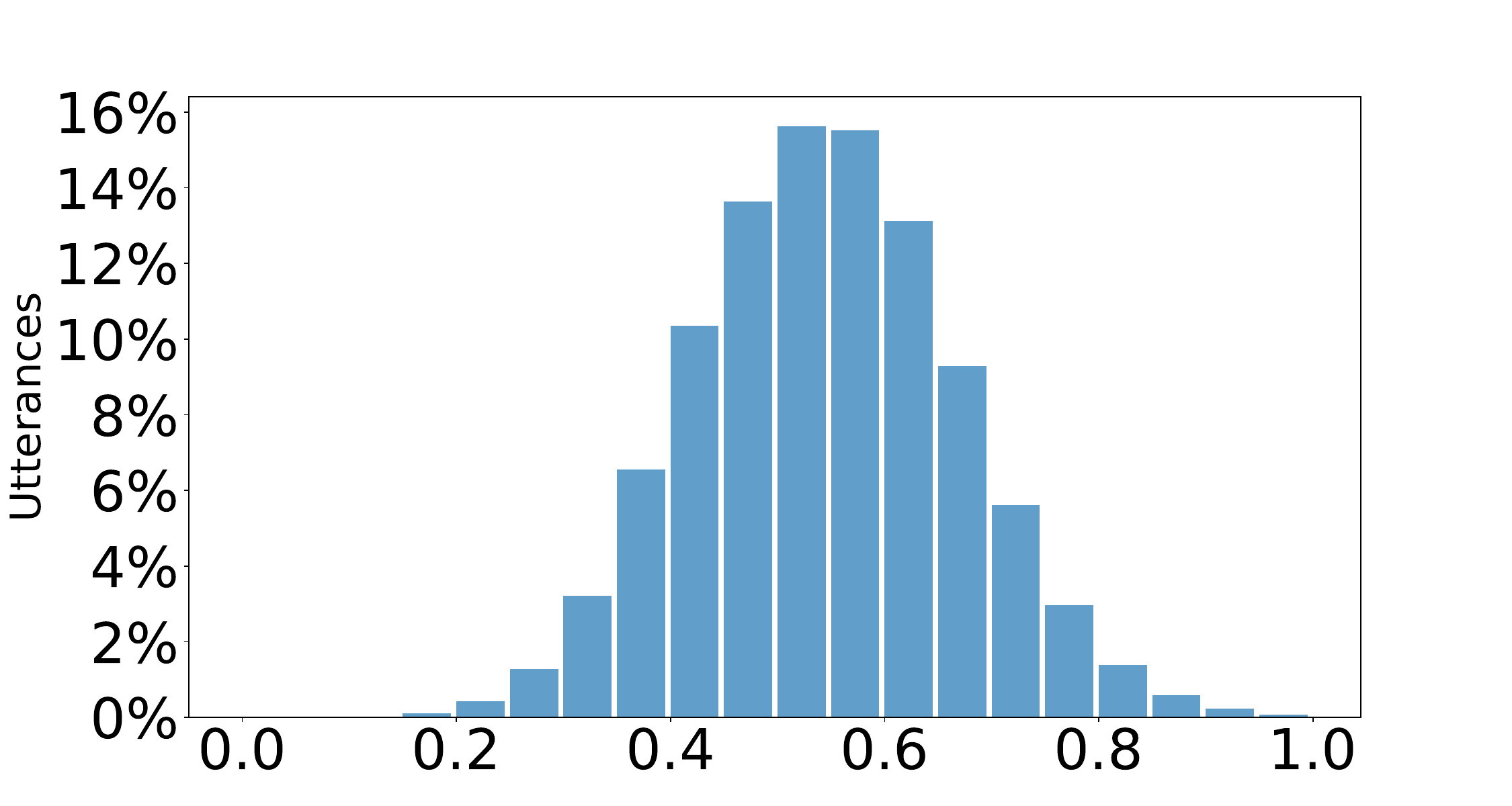}%
    } 
    \subfloat[Valence]{\includegraphics[width=0.33\textwidth]{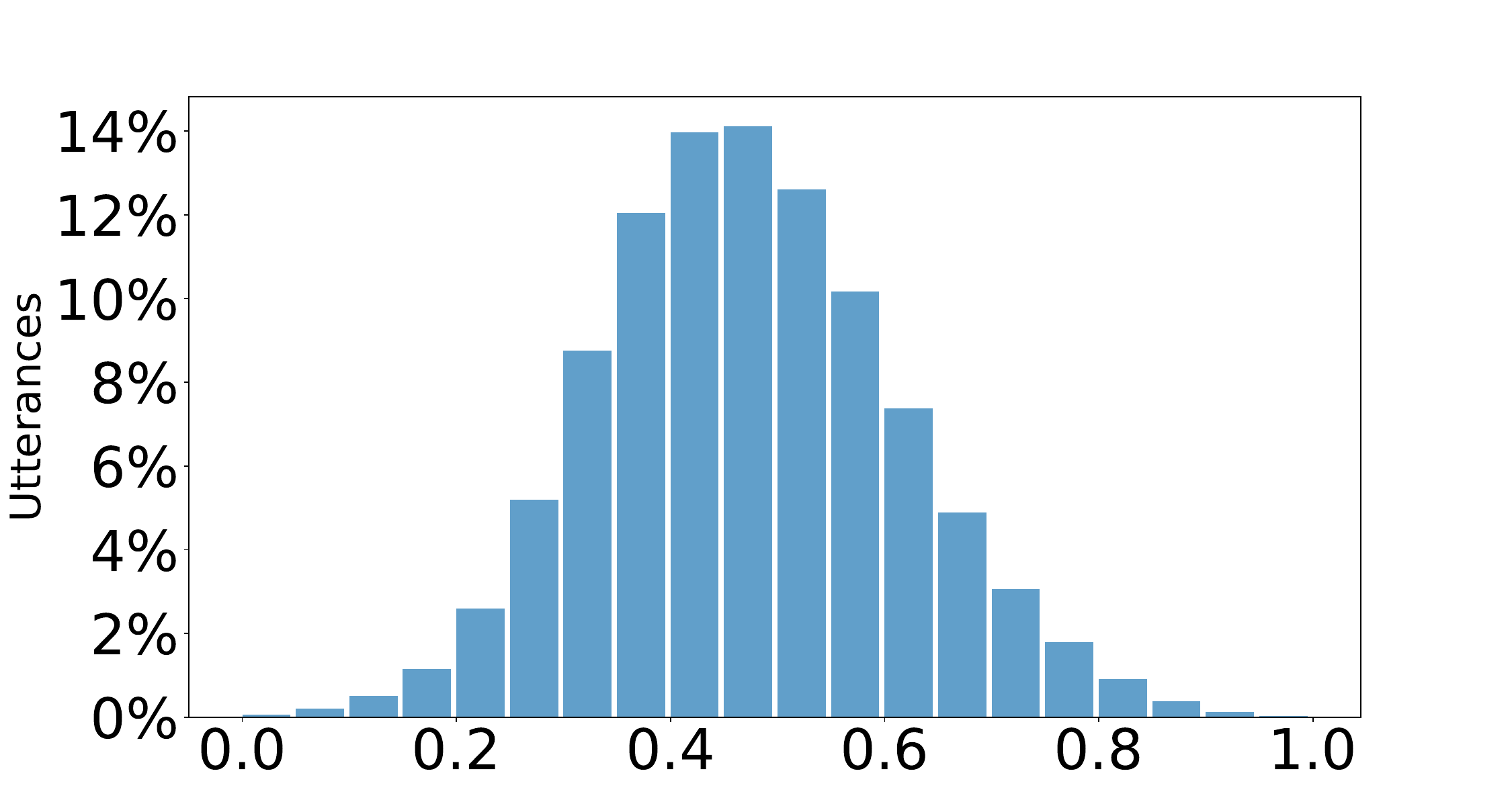}%
    }
    \caption{Distributions of continuous emotion attributes in NaturalVoices: arousal, dominance, and valence.}
    \label{fig:emotion_distribution}
    \vspace{-7mm}
\end{figure*}

\subsubsection{Text Sentiment}
We apply sentiment analysis\footnote{\url{https://huggingface.co/michellejieli/emotion_text_classifier}} to the transcriptions, classifying utterances as positive, neutral, or negative. As shown in Figure~\ref{fig:Spoken_Content_Analysis}(b), most utterances are neutral (57.94\%), while 18.89\% are positive and 23.18\% are negative. This distribution indicates that NaturalVoices captures diverse emotional tendencies in its linguistic content.  

\subsubsection{Emotion Categories}
Figure~\ref{fig:emotion_ca} shows the distribution of emotion categories in NaturalVoices. Neutral utterances account for the majority (59.0\%), while happiness (21.18\%), sadness (8.92\%), and anger (10.89\%) are also well represented. This distribution highlights the dataset’s emotional diversity.

\subsubsection{Emotion Attributes}
A distinctive feature of NaturalVoices is the inclusion of continuous emotion attributes, arousal, dominance, and valence that are rarely available in other voice conversion datasets. Figure~\ref{fig:emotion_distribution} shows their distributions across utterances. All three follow approximately normal curves, indicating balanced coverage of emotional expression. These continuous representations enable more nuanced modeling of emotion beyond discrete categories, supporting the development of emotionally rich VC models.

\begin{figure*}[!t]
    \centering
    \subfloat[SNR]{\includegraphics[width=0.33\textwidth]{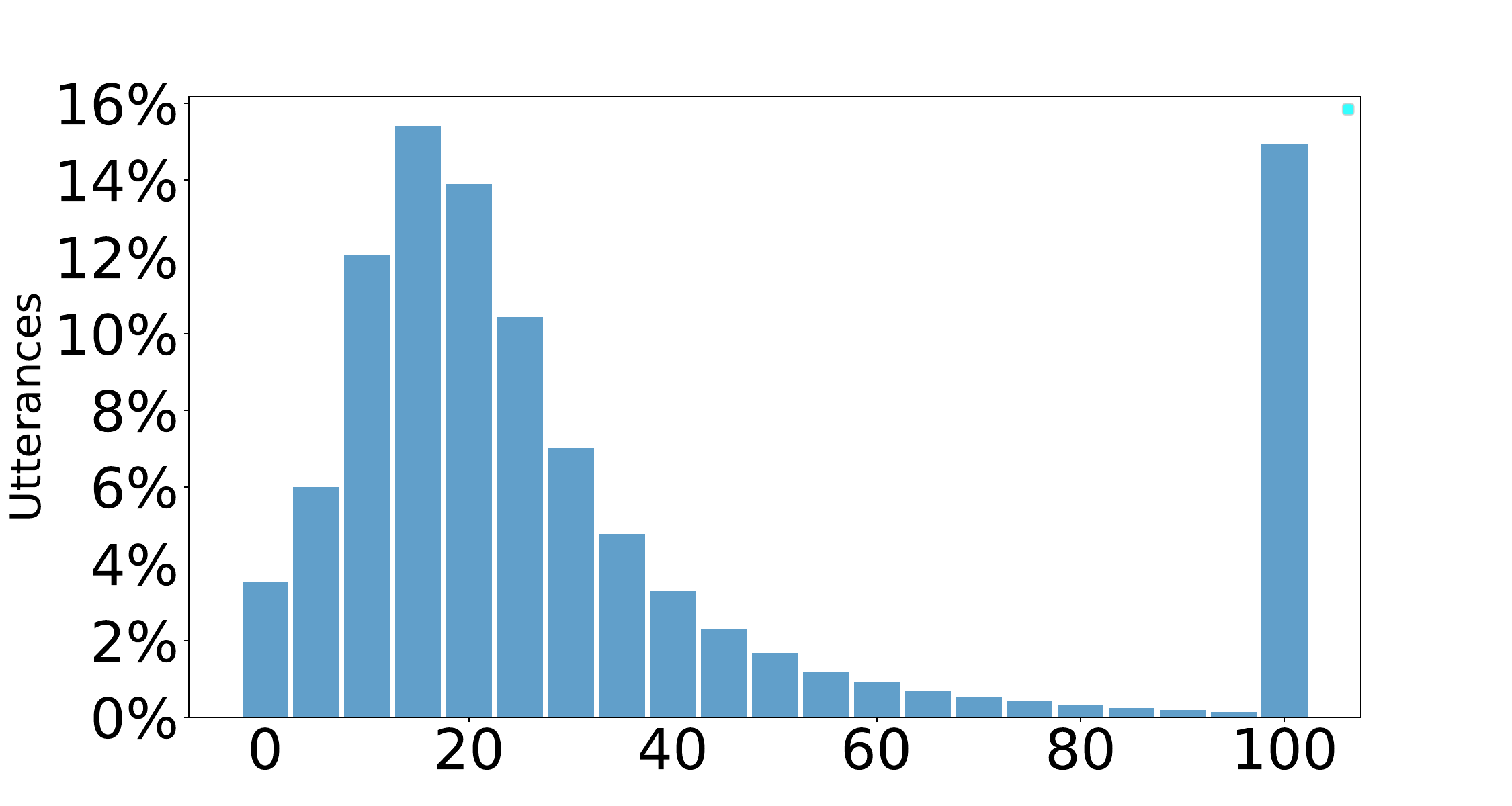}}
      \subfloat[DNSMOS Pro]{\includegraphics[width=0.33\textwidth]{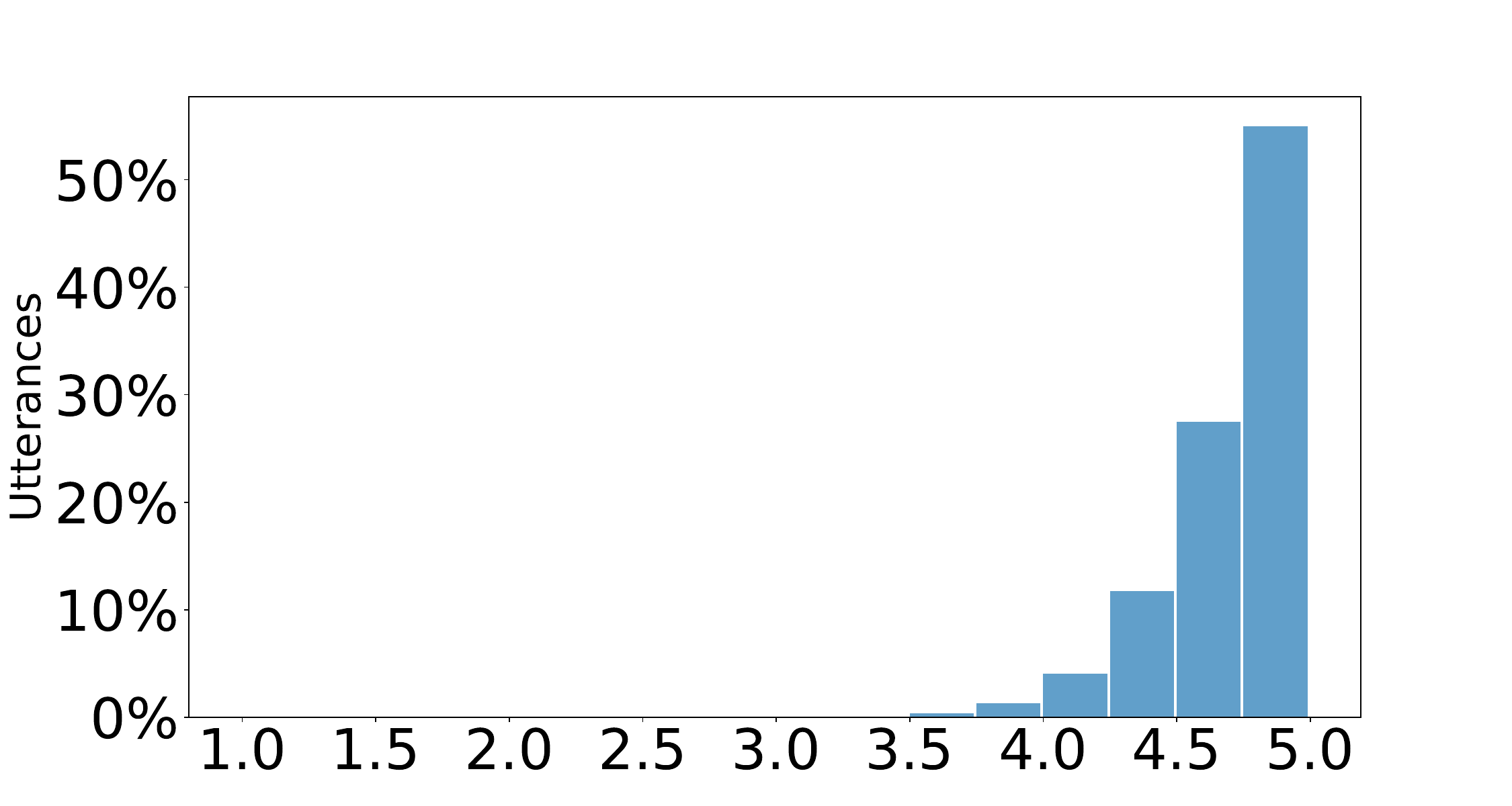}}
       \subfloat[MOS]{\includegraphics[width=0.33\textwidth]{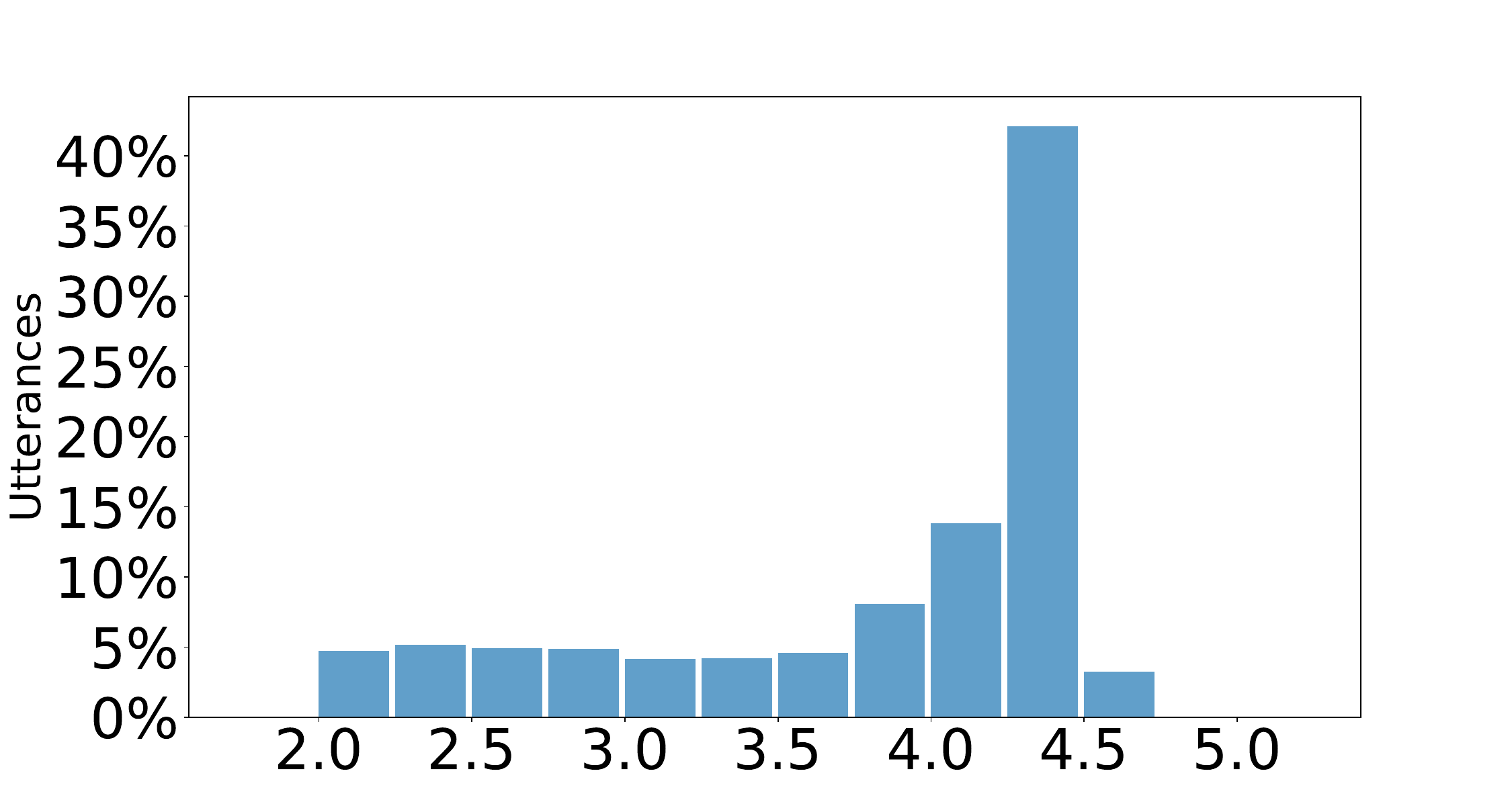}}\\
        \subfloat[PESQ]{\includegraphics[width=0.33\textwidth]{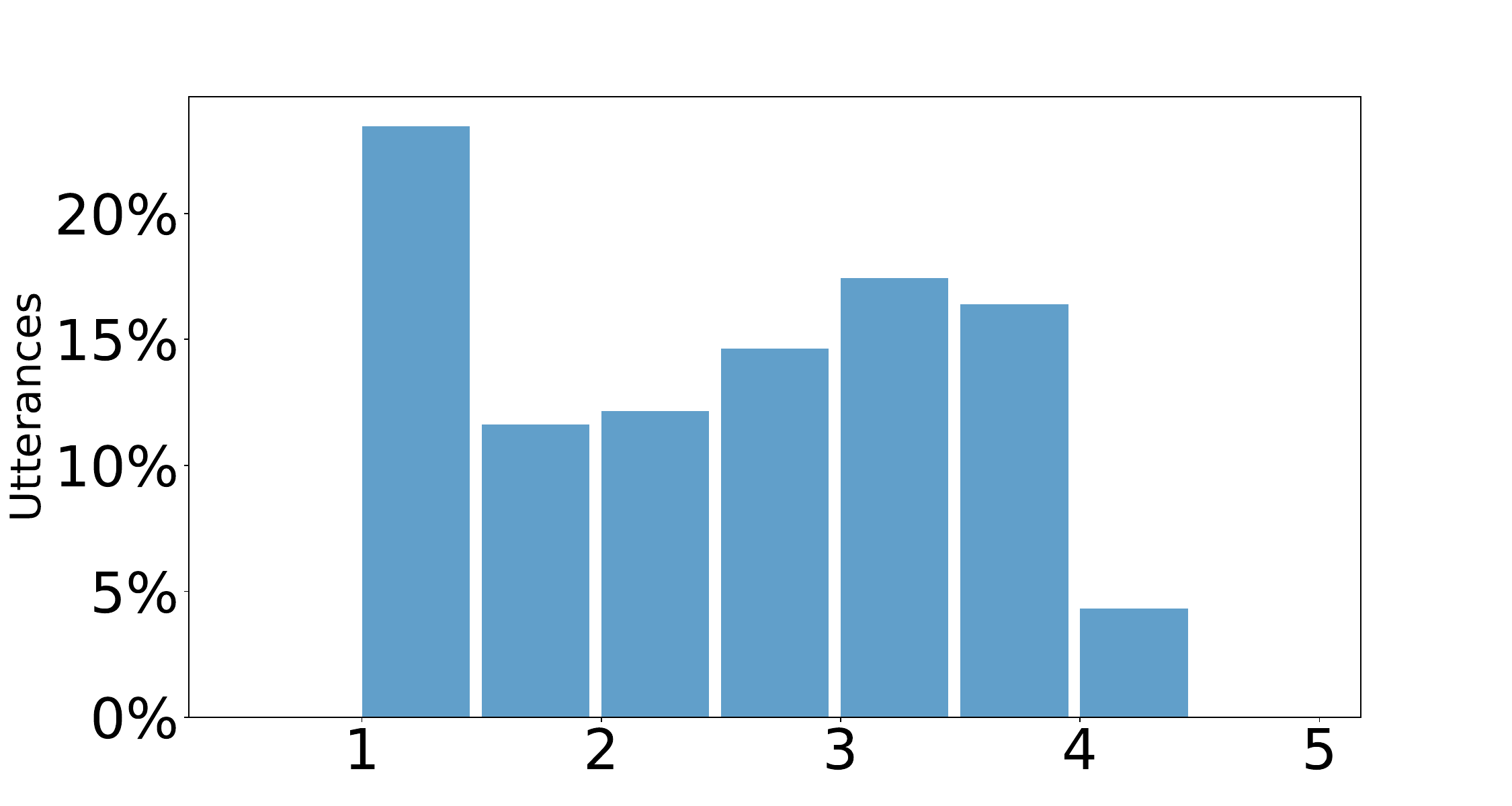}}
         \subfloat[SI-SDR]{\includegraphics[width=0.33\textwidth]{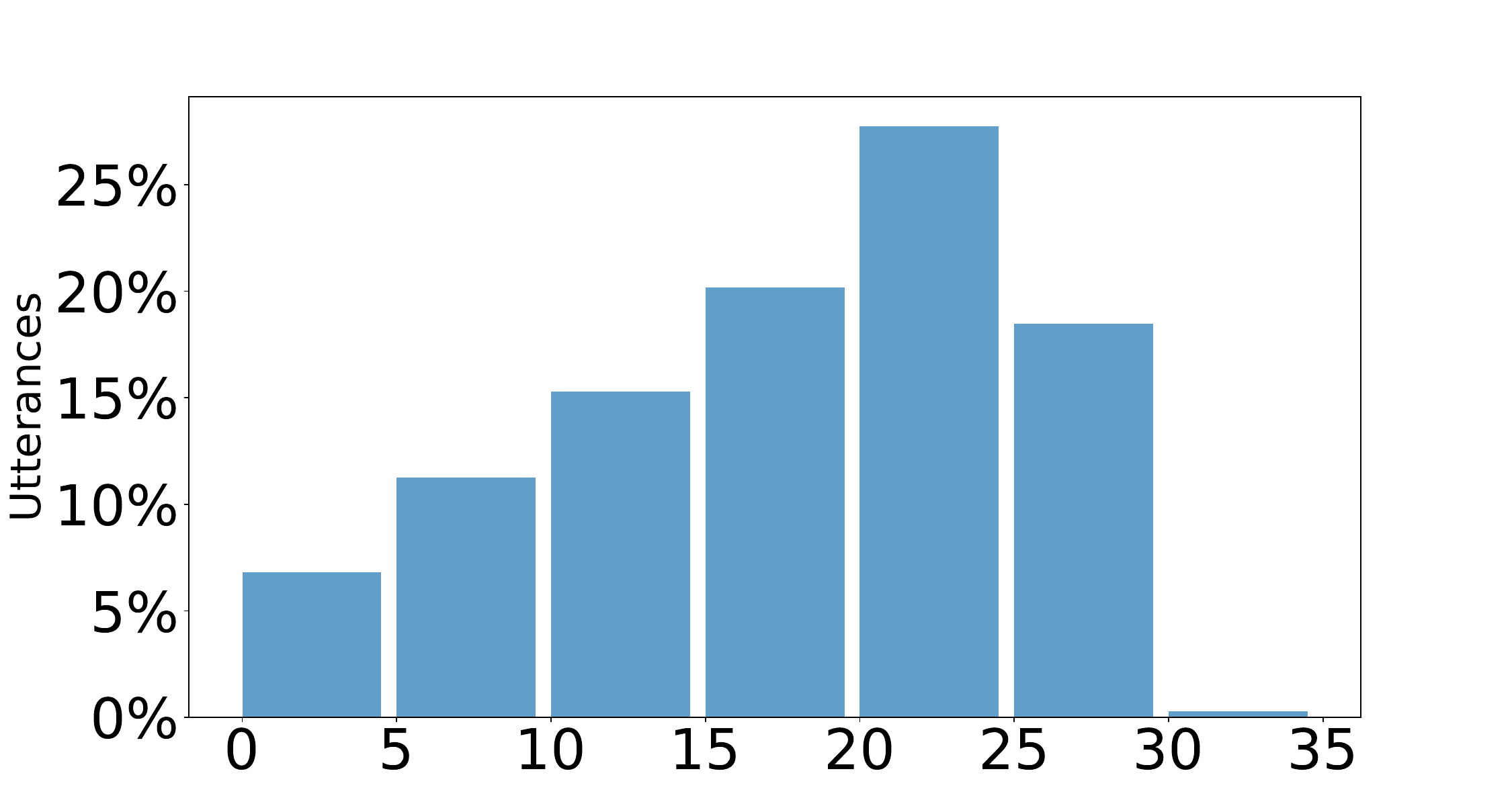}}
         \subfloat[STOI]{\includegraphics[width=0.33\textwidth]{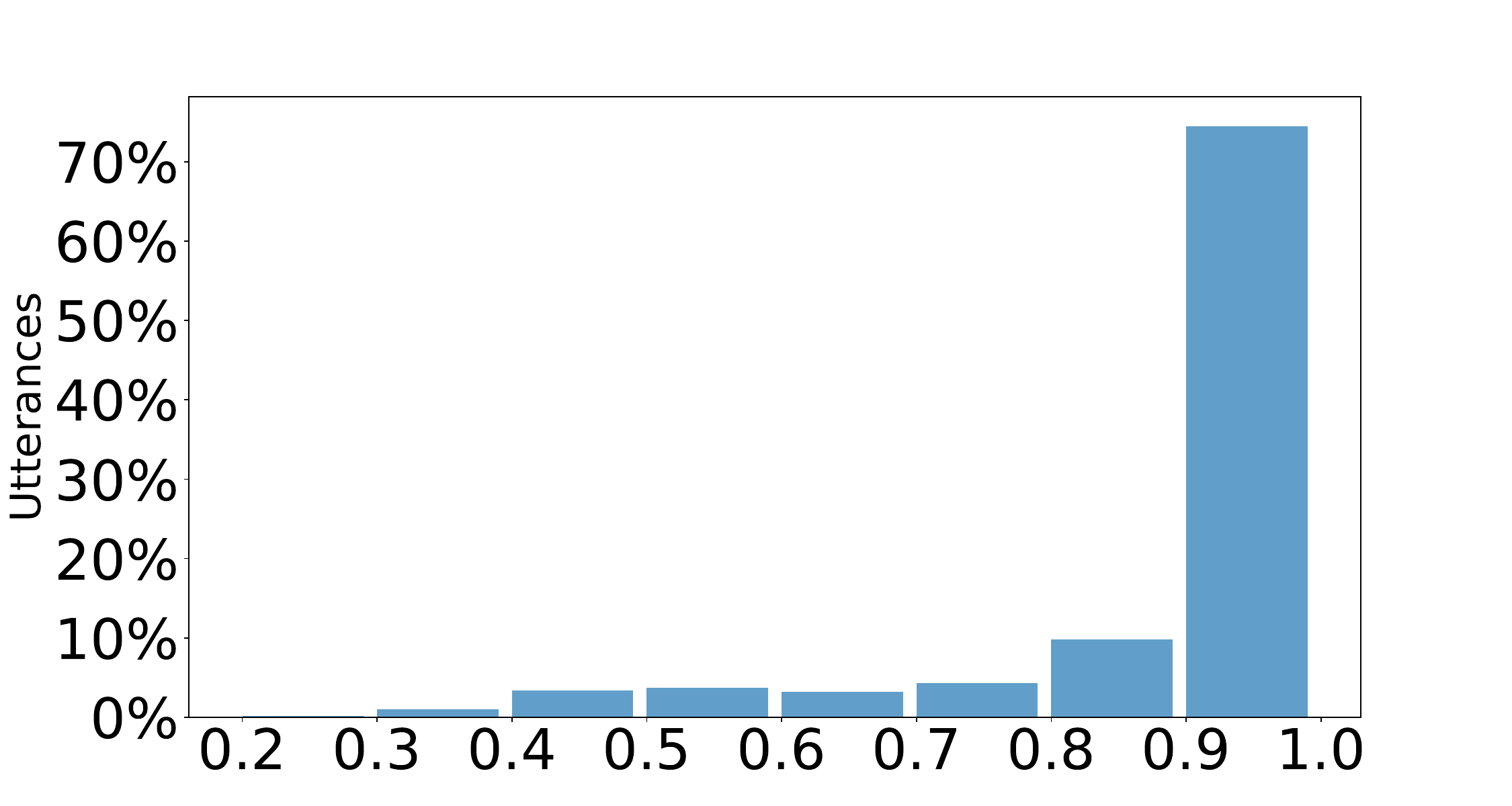}}
    \vspace{-2mm}
    
    \caption{Distributions of speech quality metrics in NaturalVoices: SNR, DNSMOS Pro, MOS, PESQ, SI-SDR, and STOI.}
    \label{fig:quality_distribution}
    \vspace{-5mm}
\end{figure*}

\begin{figure}[t]
\centering
\includegraphics[,clip,width=0.35\textwidth]{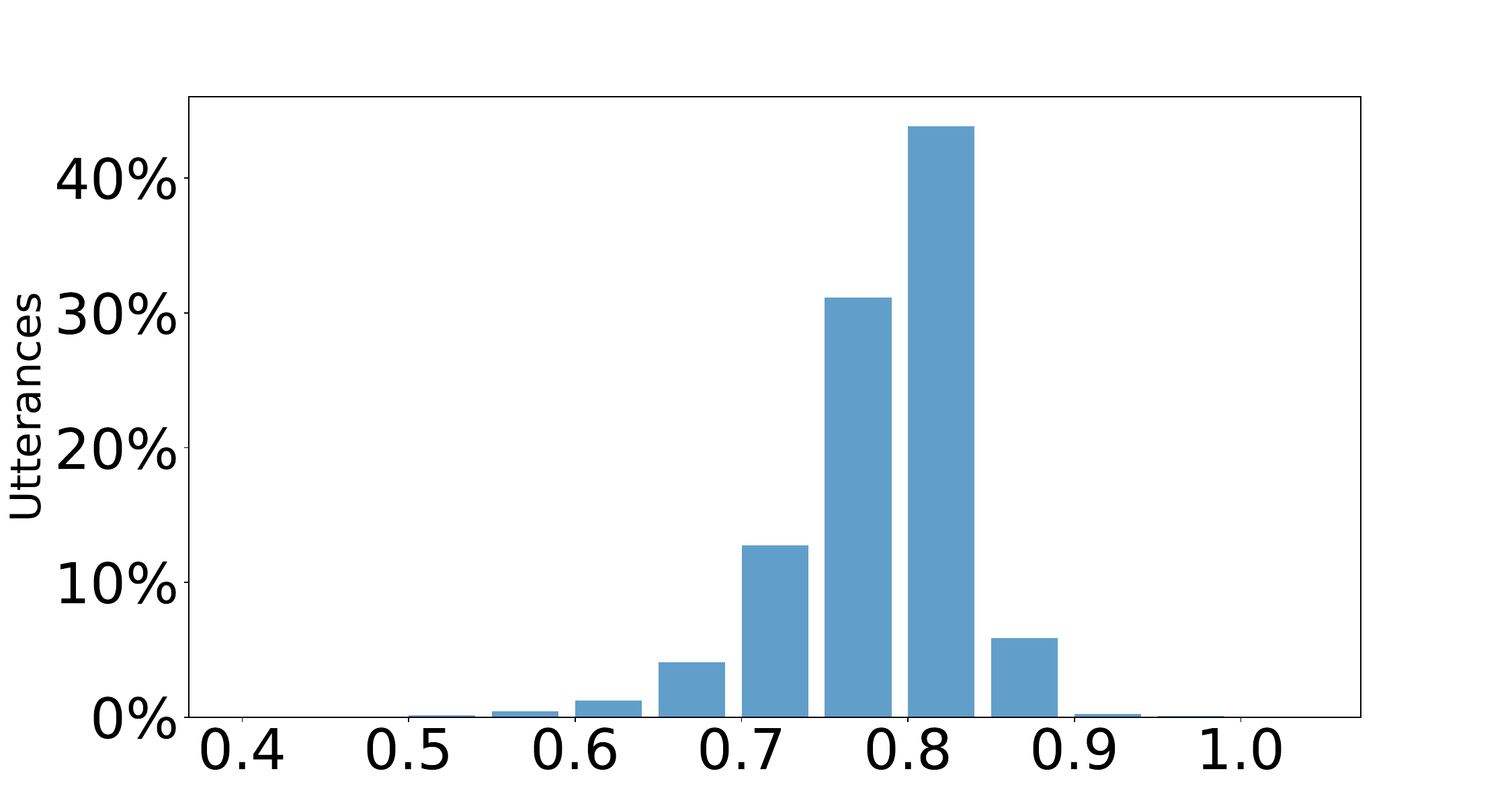}
\vspace{-3mm}
\caption{Distribution of ASR confidence scores in NaturalVoices.}
\label{fig:asr_confidence}
\end{figure}
\vspace{-5mm}
\subsection{Speech Quality Characteristics} 

We evaluate the speech quality of NaturalVoices using seven widely adopted metrics, grouped into three dimensions: noise levels, perceived quality, and intelligibility. The results are shown in Figure~\ref{fig:quality_distribution}.  

\subsubsection{Noise Levels}  
Most utterances have SNR values between 10–30 dB (Figure~\ref{fig:quality_distribution}(a)), indicating moderate to high signal clarity. A small number of segments reach extremely high SNR values (around 100 dB), reflecting near-silent backgrounds and exceptionally clean recordings.  

\subsubsection{Perceived Quality}  
DNSMOS scores are mostly above 4 (Figure~\ref{fig:quality_distribution}(b)), consistent with acceptable listening quality under typical in-the-wild conditions. MOS scores cluster between 4 and 4.5 (Figure~\ref{fig:quality_distribution}(c)), suggesting that most samples are rated as good to excellent. PESQ scores show a wider spread (Figure~\ref{fig:quality_distribution}(d)), highlighting variability in perceived quality across the dataset. SI-SDR values are generally high (around 20 dB; Figure~\ref{fig:quality_distribution}(e)), indicating low distortion and strong signal preservation.  

\subsubsection{Intelligibility}  
STOI scores remain close to 1 (Figure~\ref{fig:quality_distribution}(f)), showing that most utterances are highly intelligible. ASR confidence scores (Figure~\ref{fig:asr_confidence}) typically range from 0.7 to 0.9, confirming that the speech content is clear and transcriptions are reliable.  

Overall, NaturalVoices spans a wide range of audio quality levels. The majority of samples are well suited for voice conversion, while the variability preserves the diversity and realism of in-the-wild data.  
\subsection{Sound Events}
NaturalVoices captures a broad range of audio events typical of spontaneous, real-world speech. This diversity enhances its value for modeling and generating both speech and non-speech events in naturalistic speech synthesis.  

\subsubsection{Speech vs. Non-Speech Events}
\begin{figure}[t]
\centering
\includegraphics[,clip,width=0.35\textwidth]{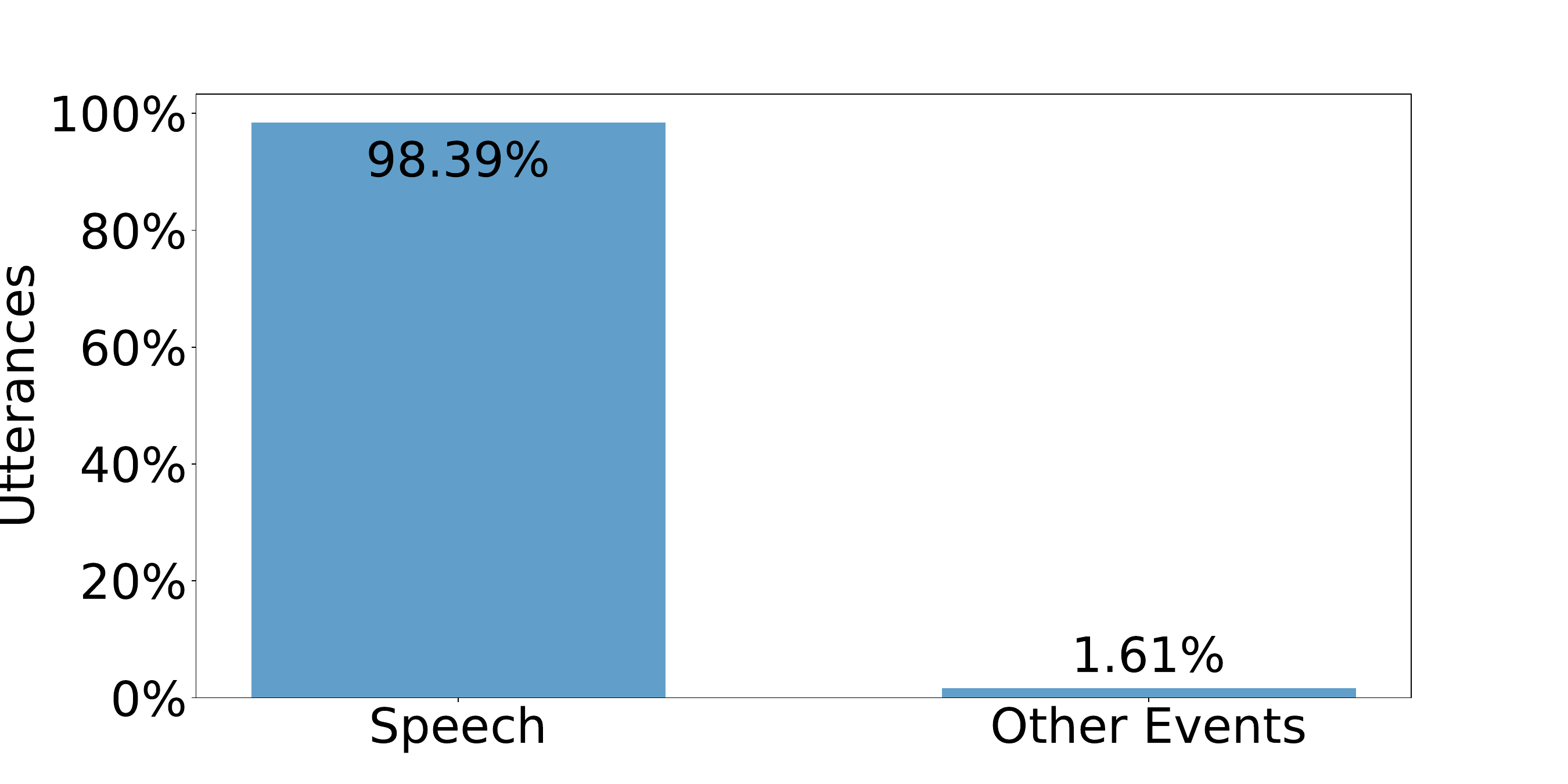}
\caption{Distribution of speech and non-speech events in NaturalVoices.}
\label{fig:sound_event}
\end{figure}

As shown in Figure~\ref{fig:sound_event}, speech accounts for 98.39\% of the dataset, while 1.61\% consists of non-speech events such as music, throat clearing, or animal sounds. This high level of speech purity makes the dataset well suited for speech processing tasks, while the presence of occasional non-speech sounds adds realism reflective of in-the-wild conversations.  
\subsubsection{Non-Speech Event Types and Durations}
Table~\ref{table:event_duration} summarize the major non-speech events and their durations. Music is the most frequent, comprising 76.1\% of all non-speech events and totaling 51.69 hours. Other events include throat clearing (2.7\%, 1.62 hours), hoots (1.7\%, 1.17 hours), and smaller contributions from gasps, sighs, and grunts. This variety enhances the dataset’s authenticity and supports applications such as expressive synthesis of nonverbal vocalizations (e.g., sighs, throat clearing).  

\subsection{Speakers per Segment}
NaturalVoices is built from podcast recordings, which typically involve multiple speakers engaged in conversational speech. As described in the previous section, these recordings are segmented into shorter utterances. We analyze the number of speakers within each audio segment to better understand the conversational structure of our dataset. 

As shown in Table IV, 4367.29 hours of audio consist of single-speaker segments, making them well-suited for traditional TTS and VC tasks. In addition, the dataset contains over 1,400 hours of multi-speaker segments: 983.13 hours with two speakers, 302.32 hours with three speakers, and 150.73 hours with more than three speakers. These segments often include overlapping speech and rapid turn-taking, offering valuable data for modeling in-the-wild, spontaneous speech. They support advanced applications such as dialogue-style voice conversion and TTS, enabling natural transitions between speakers and the synthesis of speech with subtle overlaps. This speaker diversity also facilitates the development of more context-aware and robust models, including fine-grained speaker embeddings and speaker diarization systems, which are critical for multi-speaker scenarios.

\begin{table}[h!]
\centering
\caption{Total duration (in hours) of the 9 most frequent non-speech audio events in NaturalVoices.}

\resizebox{0.33\textwidth}{!}{%
\begin{tabular}{l|c}
\hline
\textbf{Event} & \textbf{Total Duration (h)} \\ \hline
Music & 51.69 \\ 
Speech synthesizer & 2.69 \\ 
Sigh & 1.81 \\ 
Throat clearing & 1.62 \\ 
Clicking & 1.54 \\ 
Hoot & 1.17 \\ 
Frog & 1.05 \\ 
Owl & 0.95 \\ 
Hum & 0.78 \\ \hline
\end{tabular}%
}
\label{table:event_duration}
\end{table}

\begin{table}[h!]
\centering
\caption{Total duration (in hours) of audio segments in NaturalVoices, grouped by the number of speakers per segment.}
\resizebox{0.3\textwidth}{!}{%
\begin{tabular}{l|c}
\hline
\textbf{Speakers} & \textbf{Total Duration (h)} \\ \hline
1 & 2632.46 \\ 
2 & 983.13 \\ 
3 & 302.32 \\ 
More than 3 & 150.73 \\  \hline
\end{tabular}%
}
\label{table:number_speaker_duration}
\end{table}

\begin{table}[t]
    \centering
    \caption{The amount of data (in hours) used in different experiments. All subsets were randomly sampled from the 870.26-hour filtered dataset. Emo-Bal. stands for an emotion-balanced subset used at emotional VC, where each emotion category contains an equal amount of data.}
    \renewcommand{\arraystretch}{1.2} 
    \begin{tabular}{c r r r | r}
        \hline
        & \textbf{10\%}  & \textbf{50\%} & \textbf{100\%} & \textbf{Emo-Bal.} \\
       \hline
        Angry   & 8.46   & 42.26  & 89.10  & 85.00  \\
        Happy    & 11.28  & 60.89  & 124.63 & 85.00  \\
        Neutral & 55.86  & 285.24 & 571.36 & 85.00  \\
        Sad     & 8.03   & 44.03  & 85.17  & 85.00  \\
       \hline
        \textbf{Total}   & \textbf{83.63}  & \textbf{432.41} & \textbf{870.26} & \textbf{340.00} \\
        \hline
    \end{tabular}

    \label{tab:training_data}
\end{table}
\subsection{Summary and Novelty of NaturalVoices}
NaturalVoices is a large-scale podcast dataset that captures spontaneous, in-the-wild expressive speech with rich emotional attributes, balanced gender representation, diverse speech quality, and realistic conversational dynamics. In contrast to existing VC datasets that are acted, narrowly scoped, or limited in emotional diversity, it offers a naturally occurring and comprehensive resource for real-world speech applications. These qualities make it especially useful for developing expressive, emotionally nuanced, and generalizable speech models for VC and other affective computing tasks. While other large-scale real-world speech datasets for speech generation exist (e.g., \cite{he2024emilia}), they typically emphasize audio quality during preprocessing, provide only minimal annotations, and exclude segments with challenging acoustic conditions. NaturalVoices, by retaining the natural variability of podcast speech and supplying rich multi-dimensional annotations, is uniquely valuable for diverse VC tasks, including emotion-related applications. 

The inclusion of detailed speaker and emotion labels directly addresses limitations present in many existing resources, enabling the development of expressive, emotional and generalizable speech models for both voice conversion and affective computing.  

\section{Experiments on Voice Conversion}
This paper introduces the NaturalVoices dataset, which we believe will have a significant impact on the fields of voice conversion and emotional voice conversion. To assess its value as a resource, we evaluate NaturalVoices using state-of-the-art VC models under both standard and emotion-aware settings. These experiments demonstrate that NaturalVoices not only supports high-quality conversion but also serves as a realistic and challenging benchmark for advancing the field. 

\begin{table*}[h]
    \centering
    \caption{Objective evaluation results for data scaling experiments across different models and training data sizes on the two test sets.}
    \renewcommand{\arraystretch}{1.2}
    \scalebox{1}{
    \begin{tabular}{c|c|c|ccc|ccc|cc}
        \hline
        \multirow{2}{*}{Test Set} & \multirow{2}{*}{\makecell{Training \\ Data Size}} & \multirow{2}{*}{Model} 
        & \multicolumn{3}{c|}{WER} & \multicolumn{3}{c|}{CER} & \multicolumn{2}{c}{Speaker Similarity} \\
        & & & WERwhis & WERw2v & Avg & CERwhis & CERw2v & Avg  & SV Acc & SECSw \\
        \hline
        \multirow{9}{*}{NaturalVoices} 
        & \multirow{3}{*}{10\%} 
        & TriAAN-VC & 0.270 & 0.354 & 0.310 & 0.250 & 0.201 & 0.230 & 0.969 & 0.581 \\
        & & DDDMVC & 0.273 & 0.348 & 0.310 & 0.247 & 0.191 & 0.219 & 0.954 & 0.665 \\
        & & ConsistencyVC & 0.334 & 0.321 & 0.327 & 0.307 & 0.176 & 0.242 & 0.969 & 0.717 \\
        \cline{2-11}
        & \multirow{3}{*}{50\%} 
        & TriAAN-VC & 0.256 & 0.338 & 0.297 & 0.236 & 0.191 & 0.214 & 0.974 & 0.667 \\
        & & DDDMVC & 0.246 & 0.318 & 0.282 & 0.150 & 0.172 & 0.161 & 0.946 & 0.666 \\
        & & ConsistencyVC & 0.330 & 0.327 & 0.329 & 0.302 & 0.181 & 0.242 & 0.973 & 0.712 \\
        \cline{2-11}
        & \multirow{3}{*}{100\%} 
        &  TriAAN-VC & 0.175 &0.336  &0.255  &0.135  &0.190  &0.162  &0.968  &0.668 \\
        & & DDDMVC & 0.387 & 0.473 & 0.430 & 0.330 & 0.279 & 0.304 & 0.933 & 0.638 \\
        & & ConsistencyVC & 0.324 & 0.328 & 0.326 & 0.296 & 0.181 & 0.239 & 0.979 & 0.715 \\
        \hline
        \multirow{9}{*}{ESD} 
        & \multirow{3}{*}{10\%} 
        & TriAAN-VC & 0.119 & 0.175 & 0.147 & 0.072 & 0.092 & 0.082 & 0.969 & 0.639 \\
        & & DDDMVC & 0.112 & 0.162 & 0.137 & 0.066 & 0.081 & 0.074 & 0.790 & 0.643 \\
        & & ConsistencyVC & 0.119 & 0.117 & 0.118 & 0.066 & 0.053 & 0.059 & 0.908 & 0.639 \\
        \cline{2-11}
        & \multirow{3}{*}{50\%} 
        & TriAAN-VC & 0.113 & 0.160 & 0.136 & 0.069 & 0.083 & 0.076 & 0.962 & 0.647 \\
        & & DDDMVC & 0.128 & 0.162 & 0.145 & 0.074 & 0.081 & 0.077 & 0.903 & 0.653 \\
        & & ConsistencyVC & 0.105 & 0.122 & 0.114 & 0.059 & 0.056 & 0.057 & 0.960 & 0.659 \\
        \cline{2-11}
        & \multirow{3}{*}{100\%} 
        &   TriAAN-VC &0.104  & 0.159 &0.131  &0.064  &0.083  &0.074  &0.983  &0.654   \\
        & & DDDMVC & 0.249 & 0.322 & 0.285 & 0.160 & 0.176 & 0.168 & 0.768 & 0.620  \\
        & & ConsistencyVC & 0.108 & 0.121 & 0.115 & 0.060 & 0.055 & 0.058 & 0.980 & 0.690 \\
        \hline
    \end{tabular}}
    \label{tab:ds_results}
\end{table*}

\begin{table}[h]
    \centering
        \caption{MOS results with 95\% confidence intervals for two test sets.}
    \renewcommand{\arraystretch}{1.1}
    \scalebox{1.2}{
    \begin{tabular}{c|cc}
        \hline
         &  NaturalVoices & ESD \\
        \hline
         
       Speech Quality &4.51$\pm$0.18 & 4.39$\pm$0.17  \\
        \hline
    \end{tabular}}
    \label{tab:results}
\end{table}
\begin{table*}[h]
    \centering
        \caption{MOS results with 95\% confidence intervals for data scaling experiments across different models and training data sizes on the two test sets.}
    \renewcommand{\arraystretch}{1.1}
    \begin{tabular}{c|c|cc|cc}
        \hline
        \multirow{2}{*}{Test Set} &\multirow{2}{*}{Model} &  \multicolumn{2}{c|}{Speech Quality} & \multicolumn{2}{c}{Speaker Similarity} \\
        & & 10\% & 100\% & 10\% & 100\% \\
        \hline
        \multirow{3}{*}{NaturalVoices} 
        & TriAAN-VC & 2.81$\pm$0.41  & 2.79$\pm$0.47  & 3.64$\pm$0.32  & 3.68$\pm$0.30 \\
        & DDDMVC & 3.26$\pm$0.41  & 2.89$\pm$0.48 & 3.31$\pm$0.47  & 3.03$\pm$0.52 \\
        & ConsistencyVC &3.83$\pm$0.40   &3.73$\pm$0.40   &3.90$\pm$0.28  &3.88$\pm$0.27  \\
        \hline
        \multirow{3}{*}{ESD} 
        & TriAAN-VC & 2.89$\pm$0.49  & 2.94$\pm$0.53  & 3.31$\pm$0.47  & 3.03$\pm$0.52 \\
        & DDDMVC &3.30$\pm$0.33  & 2.96$\pm$0.40   &3.40$\pm$0.49   &3.22$\pm$0.45  \\
        & ConsistencyVC  &3.90$\pm$0.34  &4.07$\pm$0.33   &3.79$\pm$0.56  &3.97$\pm$0.41   \\
        \hline
    \end{tabular}
    \label{tab:results}
\end{table*}

\subsection{Research Questions}

Our experiments are designed to assess NaturalVoices across multiple state-of-the-art VC models and tasks. We aim to evaluate not only how well models perform when trained on NaturalVoices, but also what these results reveal about the strengths of the dataset and the limitations of current architectures.  

Specifically, we address three central research questions:  
\begin{itemize}  
    \item RQ1: Can NaturalVoices support high-quality VC across different architectures?
    \item RQ2: Does the filtering process produce reliable subsets that enable competitive intelligibility, speaker similarity, and emotion similarity?  
    \item RQ3: How well do models trained on NaturalVoices generalize to out-of-domain datasets (e.g., trained on spontaneous speech and tested on acted emotional speech)?

\end{itemize}  

To answer these questions, we conduct two sets of experiments: (i) a data scaling experiment, where models are trained on 10\%, 50\%, and 100\% of the filtered subset (see Section \ref{sec:data_filter}), and (ii) an emotional VC experiment, where a model is trained on an emotion-balanced subset. All models are evaluated on both NaturalVoices (in-domain condition) and
ESD (acted, out of domain condition) using a combination of objective metrics and subjective listening tests.  
 \begin{figure}[t]
\centering
\includegraphics[,clip,width=0.28\textwidth]{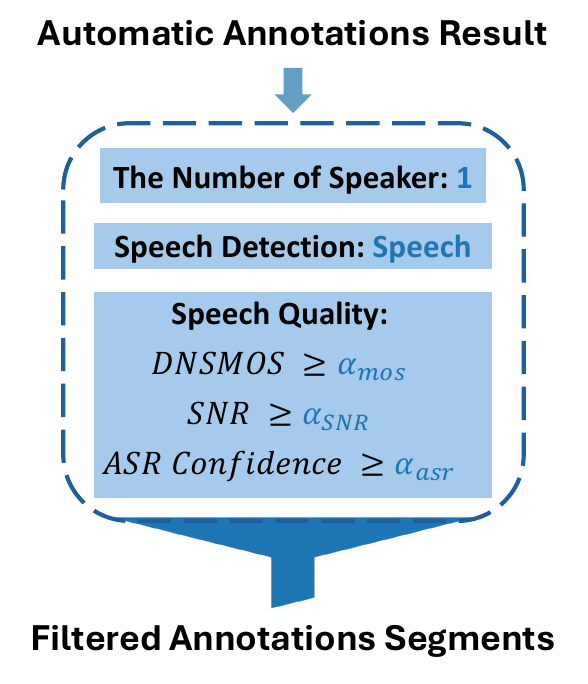}
\vspace{-3mm}
\caption{An illustration of the filtering process in our pipeline.}
  \label{fig:filtering}
\end{figure} 
\subsection{Data Filtering for Voice Conversion}
\label{sec:data_filter}
As shown in Figure~\ref{fig:filtering}, we applied a filtering process to construct a subset of NaturalVoices tailored for VC tasks. The criteria were as follows:  
\begin{itemize}
    \item Speech-only segments: Only segments labeled as ``Speech'' under the automatic speech\_classification module were retained.  
    \item Single-speaker restriction: Segments were limited to those with exactly one speaker to avoid multi-speaker scenarios.  
    \item Duration constraints: Segments with durations between 1 and 20 seconds were selected to balance variability with usability.  
    \item Speech quality thresholds: To ensure intelligible, high-quality speech, we applied cutoffs of DNSMOS $\geq$ 2.6, SNR $\geq$ 30, and ASR confidence $\geq$ 0.7, where higher values reflect better quality.  
\end{itemize}  
Applying these filters yielded 870.26 hours of speech, which serves as the training foundation for our VC experiments. This subset preserves the realism of in-the-wild speech while meeting the quality requirements for robust voice conversion modeling.  
\begin{table*}[h]
    \centering
    \caption{Objective evaluation results for emotional VC experiments on the two test sets using DISSC\cite{maimon-adi-2023-speaking} model trained on Emo-Bal subset.}
    \renewcommand{\arraystretch}{1.2}
    \scalebox{1}{
    \begin{tabular}{c|ccc|ccc|cc}
        \hline
        \multirow{2}{*}{Test Set} & \multicolumn{3}{c|}{WER} & \multicolumn{3}{c|}{CER} & \multicolumn{2}{c}{Emotion Similarity} \\
        & WERwhis & WERw2v & Avg & CERwhis & CERw2v & Avg & ECA & EECS \\
        \hline
        \multirow{1}{*}{NaturalVoices} &0.507 &0.402 &0.454 &0.445 &0.211 &0.328 &0.617 &0.724  \\
        \multirow{1}{*}{ESD} & 0.109&0.116 &0.112 &0.053 &0.049 &0.051  &0.255 & 0.286 \\
        \hline
    \end{tabular}}
    \label{tab:results}
\end{table*}

\begin{table}[h]
    \centering
    \caption{MOS results with 95\% confidence intervals for Emotional VC experiments using DISSC \cite{maimon-adi-2023-speaking} on the two test sets.}
    \renewcommand{\arraystretch}{1.2}
    \scalebox{1}{
    \begin{tabular}{cc|c|c}
        \hline
        Test Set & Model& Speech Quality & Emotion Similarity \\
        \hline
        \multirow{2}{*}{\makecell{NaturalVoices \\ (Emb-Bal)}} & GT & 4.34$\pm$0.27& - \\
        & DISSC &2.99$\pm$0.35& 3.51 $\pm$0.39 \\
        \hline
        \multirow{2}{*}{ESD} & GT & 4.27$\pm$0.28& -  \\
        & DISSC &3.82$\pm$0.28& 3.05$\pm$0.49 \\
        \hline
    \end{tabular}}

    \label{tab:results}
\end{table} 
\vspace{-3mm}
\subsection{Voice Conversion Experiments}
\subsubsection{Data Scaling}
For the data scaling experiments, we investigate how dataset size impacts VC model performance by considering three training settings. We define the 870-hour subset as the 100$\%$ data setting and randomly sample 10$\%$ and 50$\%$ of this subset for model training. The exact data distribution is presented in Table 5.{\footnote{NaturalVoices voice conversion subsets and trained models are here:\\ \url{https://huggingface.co/JHU-SmileLab}}} To assess the models' generalization ability, we constructed two test sets:
\begin{itemize}  
\item NaturalVoices (in-domain): Five male and five female speakers from the 870-hour subset, with 30 utterances per speaker. 

\item ESD Test Set (out-of-domain): Ten English speakers (5 male, 5 female) from ESD \cite{zhou2022emotional}, also with 30 utterances per speaker.  
\end{itemize}  

\subsubsection{Baselines} We benchmark three state-of-the-art any-to-any VC models:  
\begin{itemize}  
\item TriAAN-VC \cite{park2023triaan}: Uses adaptive attention normalization to enhance speaker similarity while preserving content. 
\item ConsistencyVC \cite{10389651}: Incorporates speaker consistency loss for expressive VC, aligning well with the emotional variability of NaturalVoices.  
\item DDDM-VC \cite{choi2024dddm}: A diffusion-based model with style encoding and prior mixup, designed for robust voice style transfer.  
\end{itemize}  

All models were trained on a single NVIDIA RTX 3090. Hyperparameters and checkpoints are available online.

\subsubsection{Objective Evaluations}

We assess intelligibility using Word Error Rate (WER) and Character Error Rate (CER) from two ASR systems: Whisper\footnote{\url{https://github.com/SYSTRAN/faster-whisper}} and wav2vec 2.0.\footnote{\url{https://huggingface.co/facebook/wav2vec2-large-960h-lv60-self}} We report WER\textsubscript{Whis}, WER\textsubscript{w2v}, CER\textsubscript{Whis}, CER\textsubscript{w2v}, and their averages, where lower values indicate better intelligibility. Speaker similarity is measured using (i) speaker verification accuracy with Resemblyzer \cite{wan2018generalized}, and (ii) speaker embedding cosine similarity (SECS) with Wespeaker \cite{wang2023wespeaker}, where higher values indicate stronger preservation of target identity.  

\subsubsection{Subjective Evaluations}
We conducted MOS listening tests for speech quality and speaker similarity. Twelve listeners rated 224 model-generated utterances on a 5-point scale, with 95\% confidence intervals reported. In this listening experiment, we also include the ground truth reference speech samples from both NaturalVoices and ESD to directly compare the perceived naturalness of our dataset with a well-established emotional speech corpus.

\subsubsection{Results and Discussion}

Table VI shows the objective evaluation results of our data scaling experiments. Overall, increasing the training data improves the intelligibility of the generated speech, especially for TriAAN-VC, which shows consistent gains on both NaturalVoices and ESD. ConsistencyVC remains stable across scales, while DDDM-VC shows less consistent gains and occasional degradation. Among all models, TriAAN-VC performs the best overall, achieving the lowest WER and CER when trained on 100\% of the data. While NaturalVoices is a seen test set, its real-life conditions make it more challenging than ESD, which is cleaner and recorded in studio environments. This observation highlights the importance of both data scaling and model robustness for real-world VC tasks. Speaker similarity is generally high (SV Acc 0.95–0.98) on NaturalVoices, confirming that all models capture speaker identity effectively. Increasing training data improves speaker similarity, particularly for TriAAN-VC and ConsistencyVC. However, DDDM-VC performs the best at 50\%, with speaker similarity declining at 100\%, especially on ESD (SV Acc = 0.768, SECS = 0.620). These findings suggest that current VC architectures are not yet optimized to fully leverage large-scale, in-the-wild data, underscoring the need for models explicitly designed for expressive, real-world speech.

We summarize the results of the subjective evaluation in Table~VII and Table~VIII. Table~VII compares perceived speech quality across the two test sets. NaturalVoices achieves a higher MOS than ESD, demonstrating that NaturalVoices provides natural and high-quality reference speech comparable to, and in some cases exceeding, the widely used ESD corpus. Table~VIII examines the effect of data scaling across models. On the NaturalVoices test set, performance gains from scaling are not always consistent, which likely reflects that many current VC architectures are optimized for smaller, controlled datasets and are not yet designed to fully exploit the scale and complexity of NaturalVoices. On the ESD test set, however, ConsistencyVC benefits noticeably from larger training data, while TriAAN-VC shows modest improvements and DDDM-VC's performance declines.

These results suggest that scaling effects are model-dependent and highlight NaturalVoices as a more realistic and challenging benchmark. By exposing limitations in current VC architectures, NaturalVoices creates opportunities for developing new models that are explicitly designed to leverage large-scale, expressive, and in-the-wild speech.  

\subsection{Emotional Voice Conversion Experiments}
\subsubsection{Experimental Setup}
We conducted the experiment with a 340-hour emotion-balanced subset of NaturalVoices, containing 85 hours per emotion category. From this subset, we selected five male and five female speakers, with 30 utterances per speaker per emotion as the NaturalVoices test set. We also evaluated performance on the ESD test set, as described earlier. For this experiment, we adopted DISSC \cite{maimon-adi-2023-speaking} as the baseline model, since it is specifically designed for emotional voice conversion. Because DISSC is originally an any-to-many VC model, we replaced its speaker lookup table with d-vector embeddings to enable any-to-any conversion. Prior work \cite{10447060} has shown that d-vectors also capture emotional characteristics, making this modification a viable strategy for emotional VC. All trained models and subsets used in our experiments are publicly available.\footnote{NaturalVoices EVC subset and trained models are available here: \\ \url{https://huggingface.co/JHU-SmileLab}}
\subsubsection{Objective Evaluations}
For intelligibility and emotion transfer, we adopted the same evaluation settings as in the data scaling experiments. To assess how well the converted speech preserved the target emotional state, we used two complementary metrics:  

\begin{itemize}  
    \item Emotion Category Accuracy (ECA) \cite{chen20183}: The emotion of generated speech was classified using a pre-trained model (emotion2vec) \cite{ma2023emotion2vec} and compared to the reference label. 
    \item Emotion Embedding Cosine Similarity (EECS) \cite{oh2024durflex}: Utterance-level embeddings were extracted from both generated and reference speech and the cosine similarity was calculated between them.  
\end{itemize}  
Higher values for both metrics indicate stronger preservation of emotional characteristics in the converted speech.  

\subsubsection{Subjective Evaluations} We conducted MOS listening tests to assess speech quality and emotion similarity, following the same design as in the data scaling experiments.  

\subsubsection{Results and Discussion}
Table~IX presents the objective evaluation results for emotional VC. On NaturalVoices, the model achieves higher emotion category accuracy and emotion embedding cosine similarity, showing that the generated speech closely matches the emotional expression of the reference utterances. On the ESD test set, the performance is lower, which likely reflects a mismatch between acted emotions in ESD and the spontaneous, realistic emotions in NaturalVoices. This result highlights the importance of training and evaluating on naturalistic data. In terms of intelligibility, WER and CER are higher on NaturalVoices than on ESD, underscoring that spontaneous, in-the-wild speech presents challenges not fully addressed by current EVC architectures. This result points to a broader limitation of existing models, which have largely been optimized for smaller and more controlled datasets rather than for large-scale expressive and spontaneous corpora such as NaturalVoices.

Table~X reports subjective evaluation results. Ground-truth speech achieves high quality on both datasets, while converted speech lags behind, particularly on NaturalVoices. Interestingly, the model produces higher speech quality on ESD but stronger emotion similarity on NaturalVoices. This result indicates that while NaturalVoices is more demanding for speech quality modeling, its rich emotional variation makes it especially effective for advancing emotion transfer in VC.  
\vspace{-3mm}
\subsection{Summary}

We observe that models trained on NaturalVoices achieve higher speaker similarity and more accurate emotion transfer when evaluated on real-world expressive speech compared to speech with acted emotions. This finding reveals a clear performance gap between spontaneous, in-the-wild speech and acted corpora, highlighting the importance of using spontaneous human speech in model development to ensure better alignment with natural expressive behavior. 
Furthermore, our results show that some of the current state-of-the-art VC models, originally trained and tuned on curated datasets, fail to maintain the same speech quality when trained on spontaneous corpora. Training and evaluating with NaturalVoices exposes this discrepancy and points to an important research gap in the speech synthesis community. We believe that NaturalVoices represents a high-quality and impactful resource for voice conversion that opens new directions for developing robust, expressive, and generalizable VC systems.

\vspace{-2mm}
\section{Possible Applications of NaturalVoices}
Beyond VC, NaturalVoices enables new research directions in speech processing. Its scale, speaker diversity, and real-world conditions make it valuable for tasks such as speech generation, anti-spoofing, enhancement, and speaker verification. The following subsections highlight its potential in these areas.
\vspace{-3mm}
\subsection{Speech Generation} 
Beyond VC, NaturalVoices supports broader speech generation research. Its transcripts and annotations enable tailored TTS training \cite{wang2023neural,liu2024emotion,du2024cosyvoice}, while its expressive and spontaneous recordings help models capture natural prosody and variability. Conversational segments further support dialogue-style synthesis \cite{liu2024emotion,10.1145/3664647.3681697}, with turn-taking, interruptions, and backchannels. The dataset is also well-suited for text-prompt–guided generation. Building on recent advances in prompt-based voice conversion \cite{yao2024promptvc,kuan2023towards}, NaturalVoices pairs audio with rich metadata—including emotion, speaker traits, and SNR. This metadata can be leveraged to generate natural language descriptions of emotional style \cite{chandra25_interspeech}, background noise, and sound events. This metadata makes it a valuable resource for training and evaluating models conditioned on natural language descriptions.
\vspace{-5mm}
\subsection{Anti-spoofing}
Anti-spoofing research~\cite{AS_Survey} targets detection of manipulated or synthetic audio, including replay, TTS, and VC. A key limitation is the lack of large-scale, emotionally expressive datasets~\cite{ASVspoof2019,ASVspoof2021,ASVspoof2024}, leaving models vulnerable for emotion-targeted attacks~\cite{emoAS}, where expressive synthetic speech degrades performance. Humans, by contrast, often rely on expressive cues to detect fakes~\cite{humanperception-DF}.  Recent corpora such as EmoSpoof-TTS~\cite{emoAS} and EmoFake~\cite{emofake} simulate expressive attacks, but their reliance on acted speech constrains realism. NaturalVoices, with its scale, expressiveness, and rich annotations, provides a stronger foundation for generating realistic expressive spoofs. The corpus enables training and evaluation of anti-spoofing systems under diverse, human-like expressive conditions, paving the way for more robust and prosody-aware defenses. By leveraging NaturalVoices, future anti-spoofing research can develop more robust and prosody-aware models. 
\vspace{-5mm}
\subsection{Speech Enhancement}
NaturalVoices offers valuable resources for advancing speech enhancement in realistic conditions. It contains spontaneous speech with diverse emotional expressions, varying levels and types of background noise, all of which pose meaningful challenges for enhancement models. These characteristics support the development of models \cite{9413967} that aim to preserve both intelligibility and expressive quality. Emotional speech in noisy conditions is also present in some existing speech enhancement datasets\cite{9415105, 10474162}, reflecting a growing recognition of its importance. NaturalVoices can contribute to this direction by providing a large and diverse collection of such data, supporting the development of models that preserve both intelligibility and expressive quality, particularly in self-supervised or weakly supervised settings. 

\subsection{Speaker Recognition}
In in-the-wild datasets, a major challenge for speaker verification is linking the same speaker across different audio documents. While diarization models \cite{Plaquet23,Bredin23} can assign consistent labels within a document, cross-document linking remains unresolved. NaturalVoices addresses this limitation with a hybrid strategy: combining human-annotated global speaker labels from the MSP-Podcast corpus with automatic labels produced by pre-trained diarization models. This mix of clean and noisy labels creates a valuable resource for research on semi-supervised learning, noisy-label training, pseudo-labeling, and cross-document speaker linking. Furthermore, because speaker identity and emotion are closely related \cite{10447060}, the diverse speakers and rich emotional coverage of NaturalVoices make it especially well suited for studying the interaction between speaker traits and emotional expression \cite{Parthasarathy_2017_4, pappagari2020x}.  

\vspace{-3mm}

\subsection{Audio Understanding and Reasoning}
Audio reasoning consists of a wide range of tasks that involve high-level inference and contextual understanding from audio\cite{sakshi2024mmau,xie2025audio}. For speech-based reasoning in particular, it goes beyond transcription or classification to address questions such as who is speaking, how they feel, what the intent is, what the context is, and what might happen next. These tasks require modeling long-range dependencies and complex interactions within spoken content. NaturalVoices has strong potential for training audio reasoning models. Its long-form, conversational podcast recordings provide rich context for capturing speaker dynamics, emotional shifts, and discourse flow. In addition, its rich annotations, such as speaker identity, emotion labels, and background events, enable weakly-supervised and auxiliary learning for deep speech-based audio understanding.

\section{Conclusion}
We introduced NaturalVoices, the first large-scale naturalistic podcast dataset and pipeline specifically designed for expressive and emotional voice conversion. It comprises over 5,000 hours of spontaneous podcast speech and is accompanied by a multi-module pipeline for generating detailed annotations, including transcripts, speaker identities, emotion labels, speech quality metrics, and sound event tags. Our analysis shows that NaturalVoices captures expressive, emotionally diverse, and conversational speech at scale, while its rich annotations and flexible filtering make it broadly useful across speech-related tasks. We evaluated NaturalVoices on both standard VC and emotional VC tasks. Experimental results demonstrate that models trained on NaturalVoices achieve strong intelligibility, robust speaker similarity, and effective emotion transfer, while generalizing well to out-of-domain data. By exposing the limitations of current architectures on large-scale expressive speech, NaturalVoices provides not only a valuable training resource but also a challenging benchmark for future research. Future work will explore its application in TTS, affective computing, and conversational AI, enabling more robust and expressive speech generation systems.  
\section{Acknowledgment}
This work was supported by the National Science Foundation (NSF) CAREER Award IIS-2533652. The authors also thank the Johns Hopkins University Data Science and AI Institute (DSAI) for support through a faculty startup package. We thank the MSP-Podcast team for making their dataset available for this research.
\bibliographystyle{IEEEtran}
\bibliography{refs}

\vspace{-15mm}
\begin{IEEEbiography}
[{\includegraphics[width=1in,height=1.25in,clip,keepaspectratio]{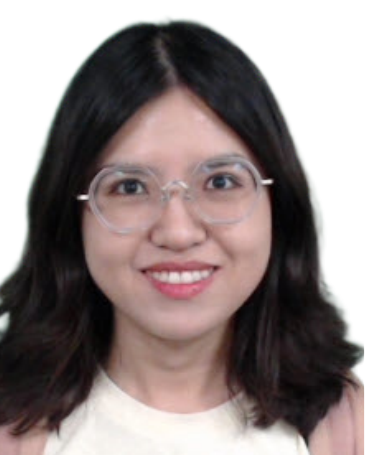}}]{Zongyang Du} (Student Member, IEEE)
received the B.S. degree in Electronic Information Engineering from the University of Electronic Science and Technology of China, Chengdu and the M.S. degree in Electrical Engineering from the National University of Singapore in 2020. She is currently pursuing a Ph.D. in Electrical Engineering at the University of Texas at Dallas and is a visiting researcher at Johns Hopkins University. Her research interests include voice conversion, speech synthesis, affective computing, and machine learning.
\end{IEEEbiography}
\vspace{-10mm}
\begin{IEEEbiography}
[{\includegraphics[width=1in,height=1.25in,clip,keepaspectratio]
{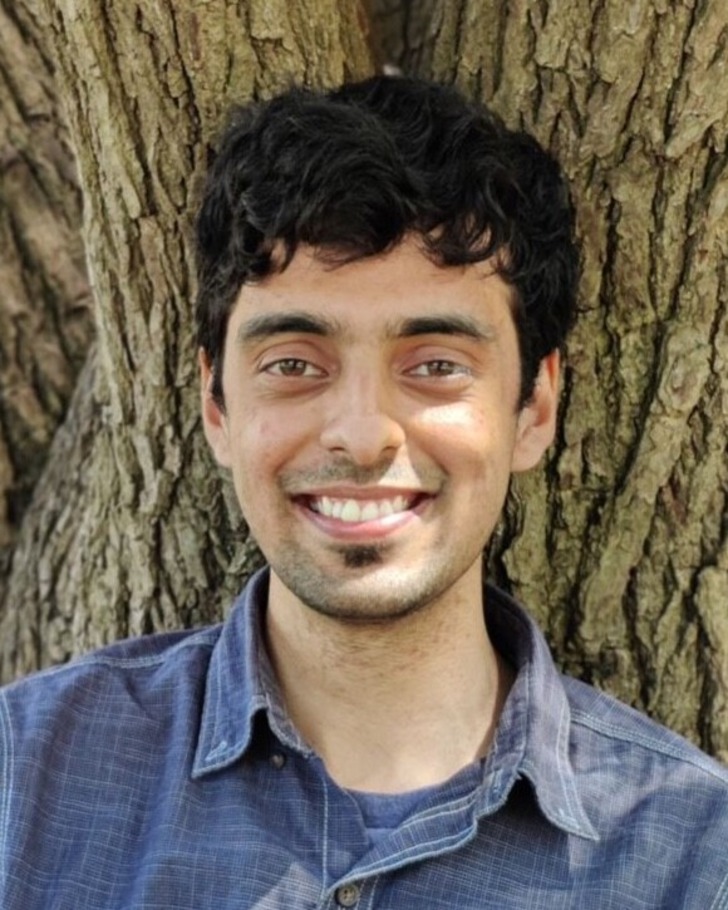}}]{Shreeram Suresh Chandra} (Student Member, IEEE) received the B.Tech degree from PES University, Bangalore, India in 2021. He is currently working towards the Ph.D. degree with Department of Electrical and Computer Engineering, The University of Texas at Dallas, USA and is a visiting researcher at the Centre for Speech and Language Processing (CLSP), Johns Hopkins University, USA. His research interests include text-to-speech synthesis, speech-language modeling, affective computing and brain-computer interfaces.
\end{IEEEbiography}
\vspace{-10mm}
\begin{IEEEbiography}
[{\includegraphics[width=1in,height=1.25in,clip,keepaspectratio]{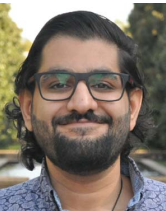}}]{Ali Salman} (Member, IEEE) received the PhD degree in electrical engineering from the University of Texas at Dallas in 2022. He received the BS and MS degrees in computer science from Indiana State University in 2015 and 2017, respectively. His current research interests include affective computing, deep learning, and facial analysis.
\end{IEEEbiography}
\vspace{-14mm}
\begin{IEEEbiography}
[{\includegraphics[width=1in,height=1.25in,clip,keepaspectratio]{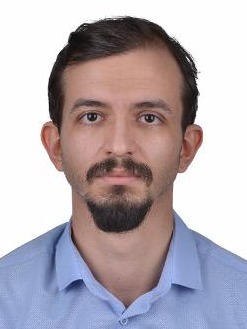}}]{Ismail Rasim Ulgen}(Student Member, IEEE) received the B.S. and M.S. degrees in Electrical and Electronics Engineering from Bogazici University, Istanbul, Turkey, in 2022. He is currently pursuing the Ph.D. degree in Electrical and Computer Engineering at Johns Hopkins University, where he is affiliated with the Center for Language and Speech Processing (CLSP). His research interests include speech synthesis, evaluation, speaker verification and emotion. 
\end{IEEEbiography}
\vspace{-12mm}
\begin{IEEEbiography}
[{\includegraphics[width=1in,height=1.25in,clip,keepaspectratio]{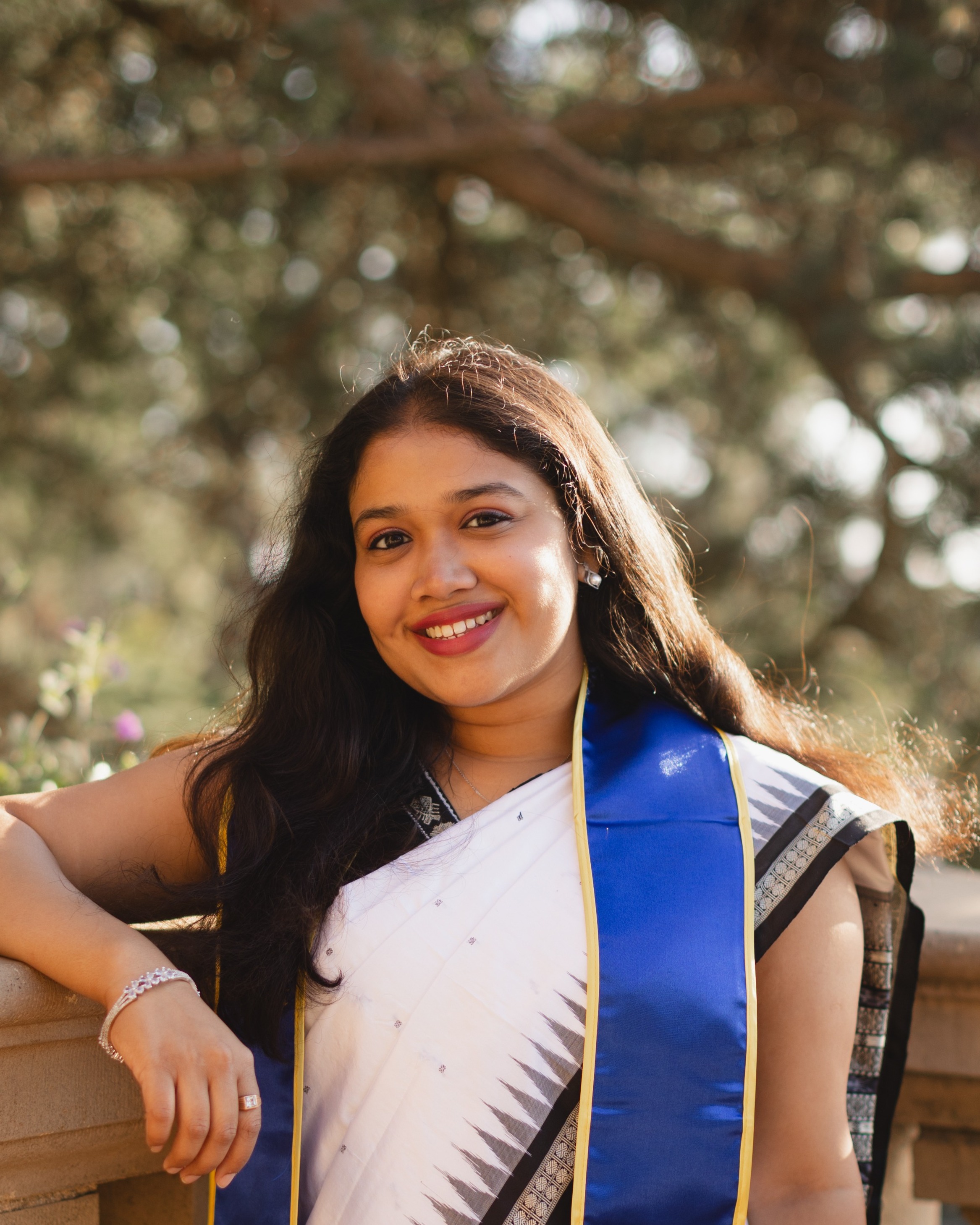}}]{Aurosweta Mahapatra} (Student Member, IEEE) received the B.Tech. degree in Electronics and Telecommunication Engineering from Kalinga Institute of Industrial Technology, Odisha, India, in 2022, and the M.S. degree in Electrical and Computer Engineering from UCLA in 2024. She is currently pursuing the Ph.D. degree in Electrical and Computer Engineering at Johns Hopkins University. Her research focuses on secure speech technologies, with an emphasis on developing robust anti-spoofing systems.
\end{IEEEbiography}
\vspace{-10mm}
\begin{IEEEbiography}
[{\includegraphics[width=1in,height=1.25in,clip,keepaspectratio]{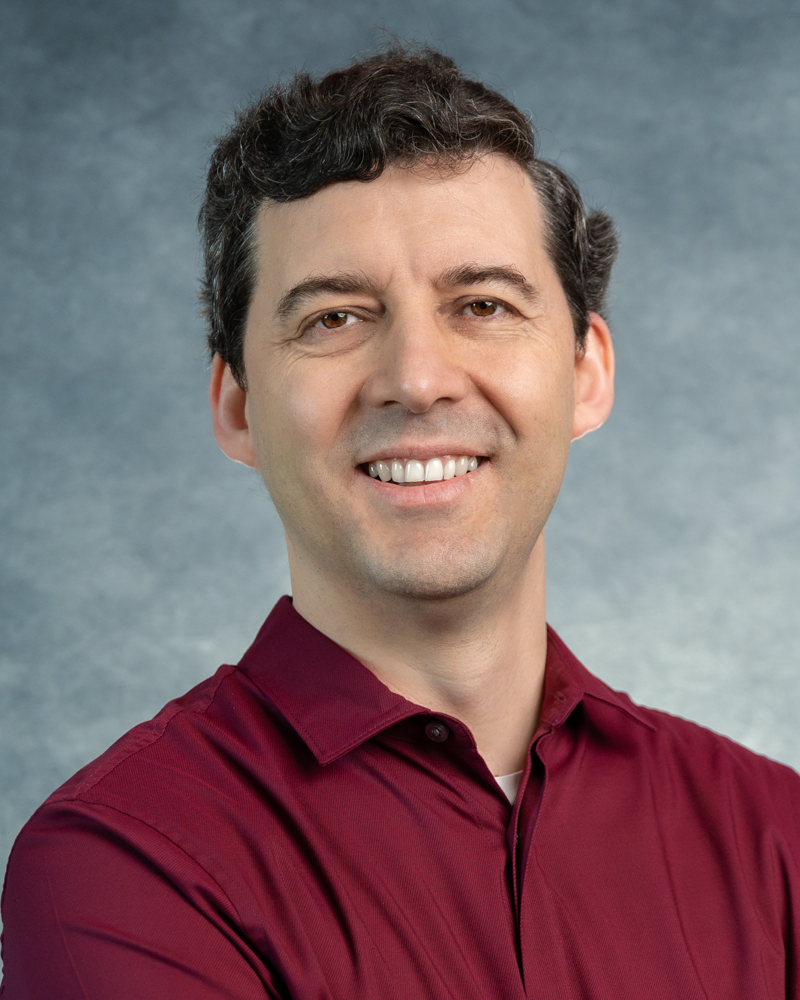}}]{Carlos Busso} (S’02-M’09-SM’13-F’23) is a Professor at Language Technologies Institute, Carnegie Mellon University, where he is also the director of the \emph{Multimodal Speech Processing} (MSP) Laboratory. He received the BS and MS degrees with high honors in electrical engineering from the University of Chile, Santiago, Chile, in 2000 and 2003, respectively, and the PhD degree (2008) in electrical engineering from the University of Southern California (USC), Los Angeles, in 2008. His research interest is in human-centered multimodal machine intelligence and applications, focusing on the broad areas of speech processing, affective computing, multimodal behavior generative models, and foundational models for multimodal processing. He was selected by the School of Engineering of Chile as the best electrical engineer who graduated in 2003 from Chilean universities. He is a recipient of an NSF CAREER Award. In 2014, he received the ICMI Ten-Year Technical Impact Award. His students received the third prize IEEE ITSS Best Dissertation Award (N. Li) in 2015, and the AAAC Student Dissertation Award (W.-C. Lin) in 2024. He also received the Hewlett Packard Best Paper Award at the IEEE ICME 2011 (with J. Jain), and the Best Paper Award at the AAAC ACII 2017 (with Yannakakis and Cowie). He received the Best of IEEE Transactions on Affective Computing Paper Collection in 2021 (with R. Lotfian) and the Best Paper Award from IEEE Transactions on Affective Computing in 2022 (with Yannakakis and Cowie). In 2023, he received the Distinguished Alumni Award in the Mid-Career/Academia category by the \emph{Signal and Image Processing Institute} (SIPI) at the University of Southern California. He received the 2023 ACM ICMI Community Service Award. He is currently a Senior Area Editor of IEEE/ACM Speech and Language Processing.  He is a member of AAAC and a senior member of ACM. He is an IEEE Fellow and an ISCA Fellow.
\end{IEEEbiography}
\vspace{-10mm}
\begin{IEEEbiography}[{\includegraphics[width=1in,height=1.25in,clip,keepaspectratio]{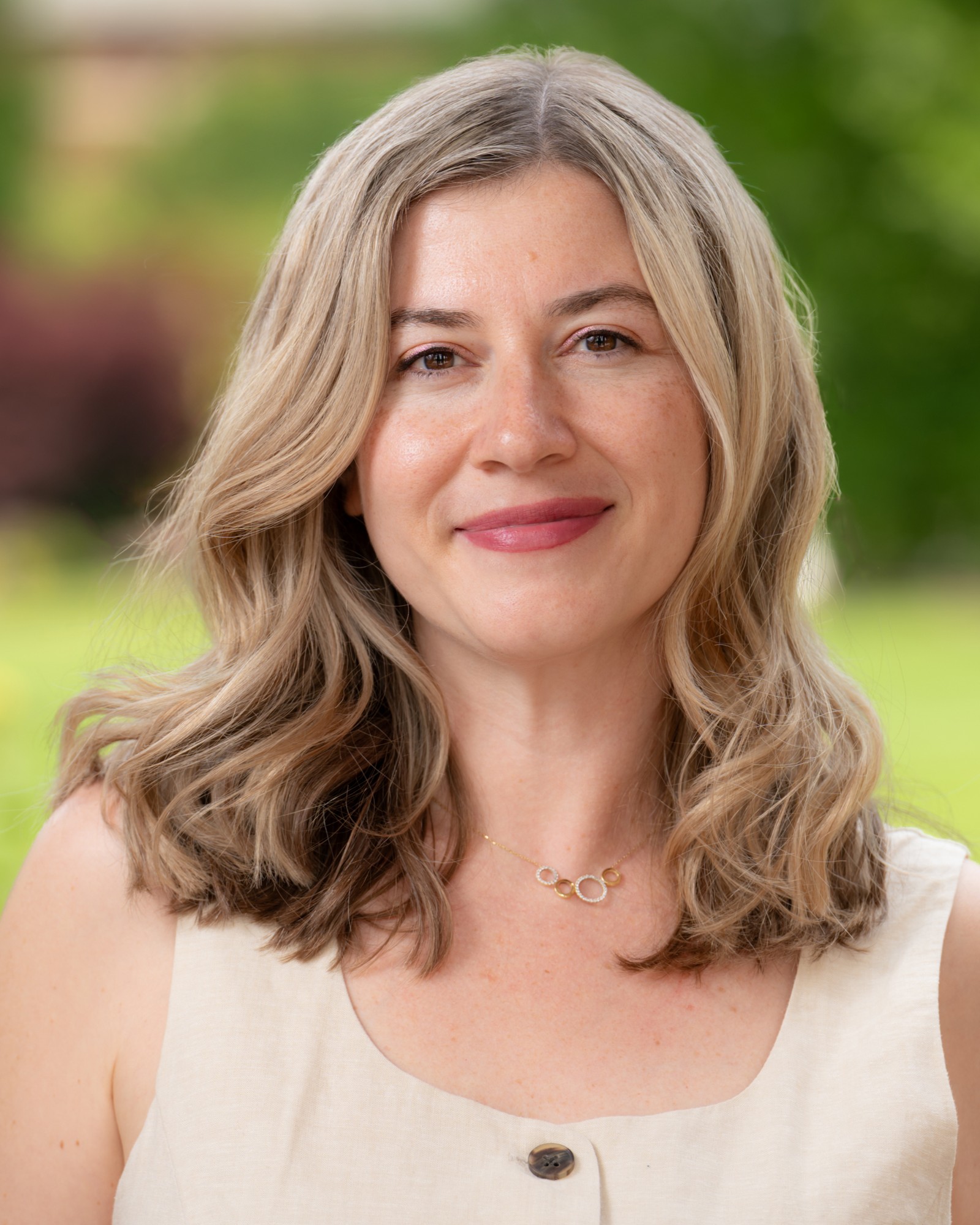}}]{Berrak Sisman} (Senior Member, IEEE) is an Assistant Professor in the Department of Electrical and Computer Engineering at Johns Hopkins University, where she is affiliated with the AI-X Bloomberg Distinguished Professorship Cluster, the Center for Language and Speech Processing (CLSP), and the Data Science and AI Institute (DSAI). She received the Ph.D. degree in Electrical and Computer Engineering from the National University of Singapore, with visiting research appointments at the University of Edinburgh, U.K., and the Nara Institute of Science and Technology (NAIST), Japan. From 2022 to 2024, she was a tenure-track faculty member at the University of Texas at Dallas. Dr. Sisman is a recipient of the NSF CAREER Award (2024), the Amazon Faculty Award (2022 and 2025), Singapore Ministry of Education Award (2021) and  A*STAR Singapore International Graduate Award (2016–2020). She is an elected member of the IEEE Speech and Language Processing Technical Committee and serves as an Associate Editor for the IEEE Transactions on Affective Computing. She is the Technical Program Co-chair for Interspeech 2026 and General Co-Chair for Interspeech 2028. She was elected to the ISCA Board for the 2025–2029 term. Her research interests include machine learning, speech processing, affective computing, speech synthesis, voice conversion and deepfake detection.
\end{IEEEbiography}
\end{document}